\documentclass{article}
\usepackage{graphicx}
\usepackage[breaklinks=false,colorlinks=true,linkcolor=blue,urlcolor=blue,citecolor=blue]{hyperref}

\usepackage{authblk} 
\usepackage[numbers]{natbib}
\usepackage{amsmath, amssymb}
\usepackage{physics} 
\usepackage{upgreek} 
\usepackage{url}

\graphicspath{{./Figures/}}

\title{First free-flight flow visualisation \\of a flapping-wing robot}

\date{}

\author[1]{Mat\v{e}j Kar\'{a}sek}
\author[1]{Mustafa Percin}
\author[2]{Torbj\o{}rn Cunis}
\author[1]{Bas W. van~Oudheusden}
\author[1]{Christophe De Wagter}
\author[1]{Bart D.W. Remes}
\author[1]{Guido C.H.E. de Croon}
\affil[1]{Faculty of Aerospace Engineering, Delft University of Technology, Kluyverweg 1, 2629 HS Delft, Netherlands}
\affil[2]{Department of Systems Control and Flight Dynamics, ONERA - The French Aerospace Lab, Centre Midi-Pyr\'{e}n\'{e}es, 31055 Toulouse, France}

\begin{document}

\maketitle

\noindent{\it E-mail}: m.karasek@tudelft.nl, m.percin@tudelft.nl, torbjoern.cunis@onera.fr, g.c.h.e.decroon@tudelft.nl

\begin{abstract}
Flow visualisations are essential to better understand the unsteady aerodynamics of flapping wing flight. The issues inherent to animal experiments, such as poor controllability and unnatural flapping when tethered, can be avoided by using robotic flyers. Such an approach holds a promise for a more systematic and repeatable methodology for flow visualisation, through a better controlled flight. Such experiments require high precision position control, however, and until now this was not possible due to the challenging flight dynamics and payload restrictions of flapping wing Micro Air Vehicles (FWMAV). Here, we present a new FWMAV-specific control approach that, by employing an external motion tracking system, achieved autonomous wind tunnel flight with a maximum root-mean-square position error of 28 mm at low speeds (0.8 - 1.2 m/s) and 75 mm at high speeds (2 - 2.4 m/s). This allowed the first free-flight flow visualisation experiments to be conducted with an FWMAV. Time-resolved stereoscopic Particle Image Velocimetry (PIV) was used to reconstruct the 3D flow patterns of the FWMAV wake. A good qualitative match was found in comparison to a tethered configuration at similar conditions, suggesting that the obtained free-flight measurements are reliable and meaningful.
\end{abstract}

\vspace{2pc}

\section{Introduction}

Flapping flight, the only form of powered aerial locomotion in nature, involves unsteady aerodynamic phenomena that remain to be fully understood, especially at small scales and low Reynolds numbers. Such understanding would be of great benefit in the development of flapping-wing Micro Air Vehicles (FWMAVs); the performance of the current designs \cite{Keennon2012,Ma2013,Gaissert2014,deCroon2009} remains far inferior compared to the extreme manoeuvrability, agility and flight efficiency of their biological counterparts \cite{Ristroph2013,Muijres2014,Sholtis2015,Bomphrey2016}. 

Despite an intense research in the fluid dynamics modelling techniques over the past decades, reliable and accurate models applicable to an arbitrary flapping wing are missing. Simpler, quasi-steady models, e.g. \cite{Dickinson1999,Sane2002,Berman2007,Armanini2016}, can successfully predict the general trends of the sub-flap forces, provided that their force coefficients are based on empirical data. Some studies capture the geometry of a deformable flapping wing during flapping, which is used as input for numerical fluid dynamics simulations \cite{Nakata2011,Deng2015,Tay2016}. While these models do provide some insight into the flow details, in most cases they still cannot predict the aerodynamic forces to a sufficient level of accuracy, as comparison to force-balance measurements reveals \cite{Deng2016thesis}. A proper numerical treatment requires coupling of models of fluid dynamics with structural dynamics of the wing in order to capture the wing deformations under the load of aerodynamic forces \cite{Nakata2011flexible}. Such models require a high computational effort while a further challenge can be an accurate identification of the structural parameters of the true wing. Thus, so far, reliable flow field data has been obtained by experimental techniques.

In biological fliers, the flow visualisation can be carried out either with tethered animals, or in-flight \cite{Thomas2004}. Tethering \cite{Ellington1996,Willmott1997c,Fuchiwaki2013,henningssonEtAl2015} typically allows for higher quality flow visualisation results, as the relative position and orientation of the animal and the measurement region can be precisely adjusted, resulting into a higher resolution data \cite{Thomas2004}. However, tethering usually also leads to unnatural wing movements so such measurements may not be representative of free flight. Therefore, there would be a strong preference to perform flow visualisation under free flight conditions.

Free flight measurements were conducted in a flight arena with hovering hummingbirds \cite{Warrick2005,Warrick2009,Altshuler2009,Pournazeri2013} and in a wind tunnel (to represent the forward flight condition) with bumblebees \cite{Bomphrey2009}, bats \cite{Hedenstrom2009,Muijres2014bats}, moths \cite{Johansson2013} or hummingbirds \cite{Ortega-Jimenez2014}. Here, the challenge is to make the animal fly at the desired position with respect to the measurement region. This typically requires intensive training and food sources, such as nectar feeders, are used to attract the animal. Nevertheless a successful measurement always requires some degree of luck due to the unpredictable behaviour of the animal. To increase the likeliness of a useful measurement, researchers typically opt for a larger measurement region, which has a trade off of lower resolution and thus less flow details captured by the measurements \cite{Thomas2004}.


Free flight experiments with flapping-wing robots would be attractive for multiple reasons. Apart from being able to quantify the effect of inherent body oscillations (present only in free-flight) on the air flow, flapping-wing robots can be programmed, meaning that the air flow could be investigated also during (controlled and reproducible) manoeuvres. Moreover, it would be possible to investigate the effect of small parameter changes, such as wing span, wing aspect ratio, etc. in a structured manner. However, until now, flow visualisation experiments with robotic flappers were conducted in a tethered condition, because precise position control necessary for successful flow measurements posed considerable challenges. Most of the studies used purposely built experimental flapping devices with model wings \cite{VanDenBerg1997a,VanDenBerg1997b,Birch2004,Lehmann2005,Truong2013,Cheng2013,Zheng2016}, while only a few works studied flight-capable FWMAVs in a tethered configuration \cite{Ren2013,Percin2014,Deng2016,Deng2016experimentalInvestigation}. 

To make the free flight flow visualisation feasible, the FWMAV needs to fly with high position accuracy. For the forward flight condition, autonomous precision wind tunnel flight has already been achieved with quadrotors \cite{wiken2015} and fixed wings \cite{nowak2010}, but FWMAVs are much more challenging to control, because of more complex dynamics and stricter weight and size restrictions on on-board computers and sensors. Our previous effort achieved the first successful autonomous wind-tunnel flight of a FWMAV \cite{deWagterEtAl2013}, but further improvements were still necessary to achieve the position and flight state stability necessary to perform such in-flight flow visualisation experiments.

\begin{figure}[ht]
\centering
\includegraphics[width=\linewidth]{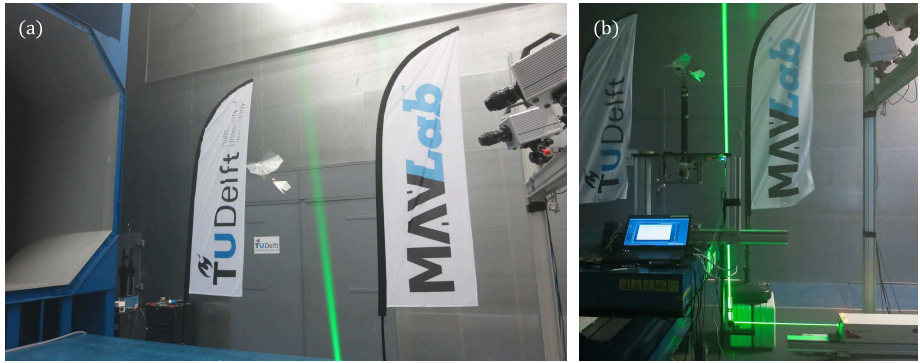}
\caption{Free flight PIV measurement of a FWMAV (a) and traditional measurement in a tethered setting used for comparison (b). A novel FWMAV-specific control approach was developed in order to achieve sufficient position accuracy and stability necessary for successful in-flight PIV measurements. The photos were taken with a reduced laser power compared to the real measurement.}
\label{fig:setup_photo}
\end{figure}

In this work, we present a methodology with which we have performed the first flow visualisation of a freely flying FWMAV (figure \ref{fig:setup_photo}a). A main component of the methodology is a novel FWMAV specific controller, which controls the MAV position in the wind tunnel, with high accuracy, through feedback from an on-board inertial measurement unit (IMU) and an external motion tracking system. In this first free-flight flow visualisation effort, a time resolved stereographic Particle Image Velocimetry (PIV) method was used to measure the wake behind the FWMAV, similar to our previous experiments with a tethered configuration \cite{Percin2014}. Thanks to the achieved control accuracy and repeatability, future analysis of flow at different locations and with different PIV methods is now possible.

In addition to the challenging free-flight PIV measurements, reference experiments were carried out under similar flight conditions, but with the same FWMAV tethered in a fixed position in the wind tunnel (figure \ref{fig:setup_photo}b). The purpose of these latter tests was to provide a comparison and assessment of the in-flight measurements, as our past study revealed differences between in-flight force estimates and clamped force balance measurements \cite{Caetano2015}. These differences, observed mainly in the direction of the stroke plane, were partly attributed to the dynamic oscillations that are present in the free flight but are restricted in the clamped measurement.



The paper is organised as follows: the experimental setup, the control algorithms and the PIV processing techniques are presented in section \ref{sec:methods}. Section \ref{sec:results} reports on the results of precision position control of the FWMAV as well as on the flow visualisation results. Finally, section \ref{sec:conclusions} provides conclusions and discusses potential future tests and improvements.

\section{Methods}
\label{sec:methods}

\subsection{Experimental setup}

\begin{figure}[h]
\centering
\includegraphics[width=\linewidth]{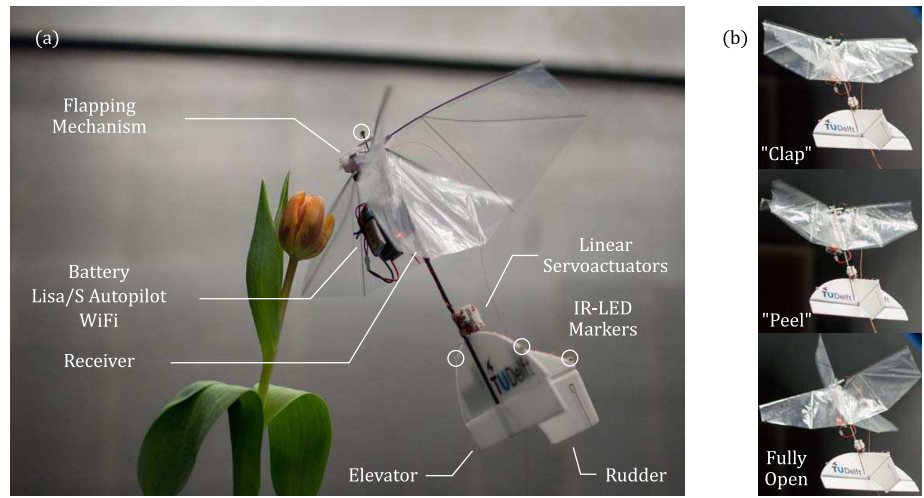}
\caption{The DelFly II FWMAV used in the tests. (a) Description of the main components. (b) The important phases of the flapping  motion, including the 'clap' and 'peel', which help enhance the lift production and efficiency. For reliable tracking, the DelFly was equipped with four active IR-LED markers, three placed on the tail and one on the nose.}
\label{fig:delfly}
\end{figure}

The experiments were carried out with the DelFly II MAV (further called simply the DelFly), a well-studied flapping-wing platform developed at TU Delft \cite{deCroon2016delflyBook}. The DelFly, displayed in figure \ref{fig:delfly} (a), is a biplane design with flexible wings (280 mm wing span) arranged in a cross configuration, moving in opposite sense while flapping. Once per wing beat cycle, the wings \textit{clap} together as they meet and \textit{peel} apart again, see figure \ref{fig:delfly} (b). This \textit{clap-and-peel} mechanism has a positive effect on the overall thrust production and efficiency \cite{Lehmann2005,DeClercq2010}. Due to its conventional tail with horizontal and vertical tail surfaces, the DelFly is inherently stable and its two control surfaces, the rudder and the elevator, are only used for steering. The DelFly has a large flight envelope, which ranges from near hover flight ($\approx$~0 m/s, vertical body orientation), to fast forward flight ($\approx$~7 m/s, nearly horizontal body orientation). For faster speeds, the centre of gravity needs to be shifted forward for inherent stability \cite{Koopmans2015}.

For the experiments described here, the DelFly was equipped with a Lisa/S autopilot board \cite{Remes2014}, which includes a 6 DOF MEMS inertial measurement unit (IMU) for on-board attitude estimation (Invensense MPU 6000) and a 72 MHz ARM CPU capable of running Paparazzi open source autopilot system \cite{Paparazzi}. The autopilot board was attached to the fuselage with a soft foam mount in order to isolate the high frequency vibration. Further components include an Mi-3A speed controller (flashed with BL heli firmware) driving the main brush-less motor (customised design with 28 turns per winding \cite{deCroon2009}), two Super Micro linear servo actuators for the tail control surfaces, a DelTang Rx31 receiver for the radio link and an ESP8266 Esp-09 WiFi module for the datalink between the autopilot and the ground station. The system was powered by a 180 mAh single cell LiPo battery (Hyperion G3 LG325-0180-1S). The overall mass of around 23 grams allowed for flight endurance between 2 to 6 minutes, depending on the flight speed.

\begin{figure}[h!]
\centering
\includegraphics[width=0.8\linewidth]{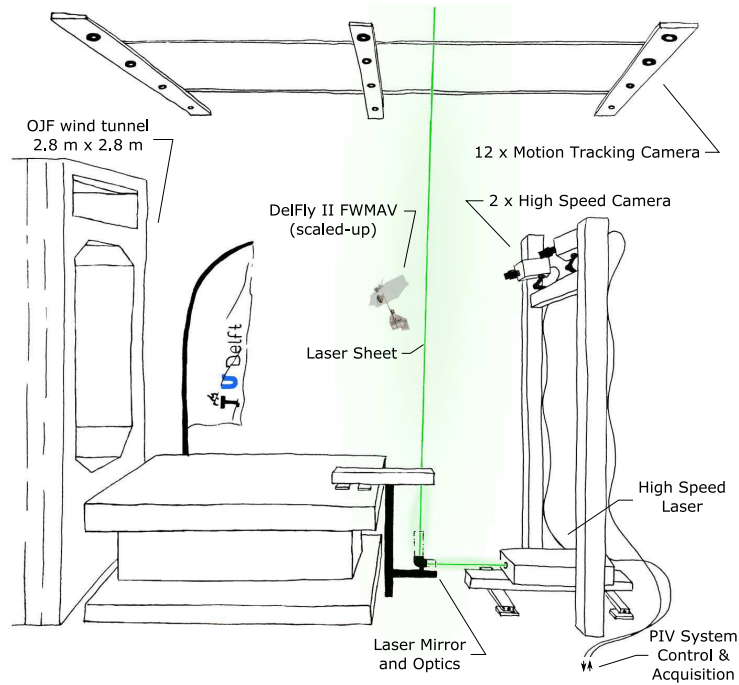}
\caption{A schematic sketch of the experimental setup. The wind tunnel room was equipped with 12 OptiTrack Flex 13 motion tracking cameras for FWMAV tracking. The stereoscopic PIV setup consisted of 2 Photron FastCam SA 1.1 high speed cameras mounted at a relative angle of around 40$^\circ$. A high speed Mesa-PIV double-pulse laser illuminated the measurement plane located about 150 mm downstream of the FWMAV tail. Its beam was expanded to form a $\approx$ 2 mm wide laser sheet. Prior to the measurements, the room was filled with water-glycol based fog of droplets in order to achieve homogenous seeding of the flow.  Illustration by Sarah Gluschitz (CC BY-ND 4.0).}
\label{fig:experimental_setup}
\end{figure}

The experiments were conducted in the Open Jet Facility wind tunnel at TU Delft, see figure \ref{fig:experimental_setup}. This return-type, low speed wind tunnel has a large open test section with a cross-section of 2.8~m $\times$ 2.8~m, providing enough space for the proposed free flight experiments. During the tests, the wind tunnel was operated at speeds ranging from 0.8~m/s to 2.4~m/s. The wind tunnel room was equipped with an OptiTrack motion tracking system (NaturalPoint, Inc.) consisting of 12 OptiTrack Flex 13 motion tracking cameras (resolution 1280~px $\times$ 1024~px, 120~fps). The system was primarily used for tracking the test aircraft position and heading, but provided also the positions and orientations of the measurement plane and the high speed cameras of the PIV system. For reliable tracking, the DelFly MAV was equipped with four active IR LED markers placed on its body according to figure \ref{fig:delfly} (a). Reflective markers were used on the remaining objects (calibration plate, PIV cameras).



The flow visualisation technique chosen for the experiments presented here is that of time-resolved stereoscopic Particle Image Velocimetry (PIV). The PIV system consists of a high speed laser and two high speed cameras which acquired images (1024 px $\times$ 1024 px) at a rate of 5 kHz. Based on our prior experience in similar experiments with a clamped FWMAV \cite{Percin2014,Deng2016}, we have opted for performing measurements in the wake behind the DelFly in order to avoid problems associated with laser reflections on the shiny surfaces of the wing and that of the wings blocking the camera view. The measurement plane was set normal to the free flow, behind the DelFly tail and an advective approach (“Taylor’s hypothesis”) was applied to reconstruct an estimate of the three dimensional wake configuration. A similar approach has been used in a variety of animal studies \cite{Hedenstrom2009,Muijres2011,Henningsson2011,Muijres2012flycatchers}.





The DelFly was controlled by the on-board autopilot, which was steering it towards a desired position set-point based on feedback from the external motion tracking system. An operator was monitoring on-line the position errors and triggered the measurement at a convenient moment. He would also repeat the measurement in case the errors were too large. Additional IR LEDs, fixed with respect to the ground and detected by the tracking system, were turned on together with the trigger signal to the PIV system, which served as a time stamp for time synchronisation of the tracking and PIV data sets. The simultaneous application of the free-flight FWMAV control and the PIV measurements required to ensure that the optical motion tracking operation was not adversely affected by the laser light and the seeding fog introduced for the PIV experiments.

\subsection{Control}

To ensure successful PIV measurements a high precision position control needs to be achieved, so that the wake of the DelFly stays within the measurement region. At the same time, because we are interested in free steady flight, the thrust and power should not vary (dramatically) during the measurement. These are two opposing requirements: the wind tunnel will always have some remaining turbulence that the controller should respond to, but if tuned too aggressively, the power will vary significantly and the controller may even respond to the inherent flapping induced body rocking.

The size of the PIV measurement region (170~mm $\times$ 170~mm) was chosen to be slightly larger than the half span of the DelFly (140 mm) so that the wake of the right half wings could be captured (a symmetry of left and right half wings was assumed). Because the dominant flow structures are observed behind the wing tips, we have estimated that a successful measurement can be carried out if the root mean square ($rms$) position error remains below 25~mm in all directions for a time course of 2 seconds (a single PIV measurement takes approximately 1 second). In order to meet these requirements, we designed a novel FWMAV-specific control scheme.

The tests presented here cover the flight speeds between 0.8~m/s and 2.4~m/s, which corresponds to body pitch angles between approximately 70$^\circ$ and 30$^\circ$, respectively. The large range of body pitch throughout the flight envelope affects the way the DelFly is controlled: in near hover flight (body almost vertical) a change of flapping frequency will affect mostly the climbing/descending while elevator deflection $\zeta$ will have a dominant effect on the body pitch and subsequently the forward speed. In fast forward flight (body nearly horizontal) the control is inverted: flapping frequency change has a dominant effect on forward speed while pitching the body through elevator deflections affects mainly climbing/descending. A rudder deflection $\eta$ will initially cause a banked turn, but will result in a pure heading change once the rudder returns back to its neutral position. This is due to a positive dihedral angle of the MAV providing inherent stability around the roll axis. Such behaviour can be observed over the whole flight envelope, but the rudder effectiveness will vary with airspeed. Thus, control of FWMAV is extremely challenging as it needs to consider all these effects.

\begin{figure}[h!]
\centering
\includegraphics[width=0.8\linewidth]{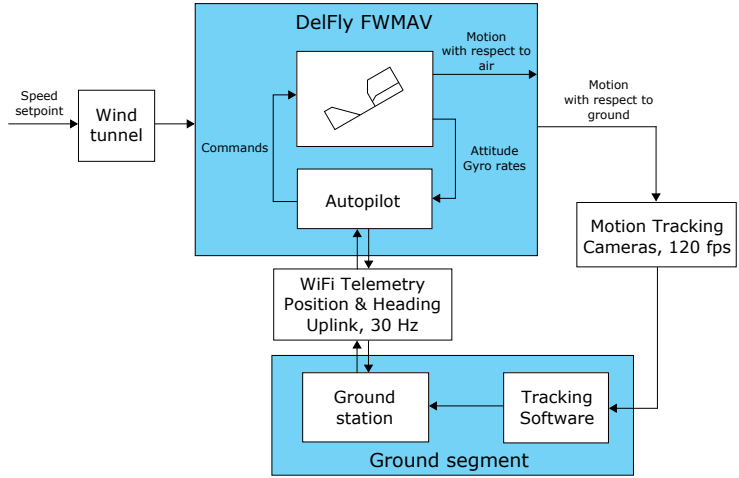}
\caption{Block diagram of the control system. The DelFly FWMAV is controlled by an on-board autopilot that uses feedback from an on-board inertial measurement unit and an external motion tracking system, which measures the FWMAV position and orientation with respect to the wind tunnel axes. A proprietary software (Motive 1.9) processes the camera data and sends the position and heading to the ground station. A WiFi uplink is used to transmit this information on-board at a rate of 30 Hz.}
\label{fig:control_overview}
\end{figure}

A general block diagram of the designed control system is in figure \ref{fig:control_overview}. The wind tunnel generates uniform airflow with a constant speed. The DelFly flies relative to the moving air and is controlled by an on-board autopilot, which steers the vehicle based on feedback from the on-board IMU (used for attitude estimation) and from an external motion tracking system that provides position and heading information (with respect to ground). The tracking system data, captured at 120 Hz, is transmitted via LAN network to the ground control station and sent further, with a rate of 30 Hz, to the autopilot using a wireless WiFi data-link. The same link is also used for telemetry that can be viewed on-line on the ground station.

\subsubsection{Axis system}
\begin{figure}[ht]
\centering
\includegraphics[width=\linewidth]{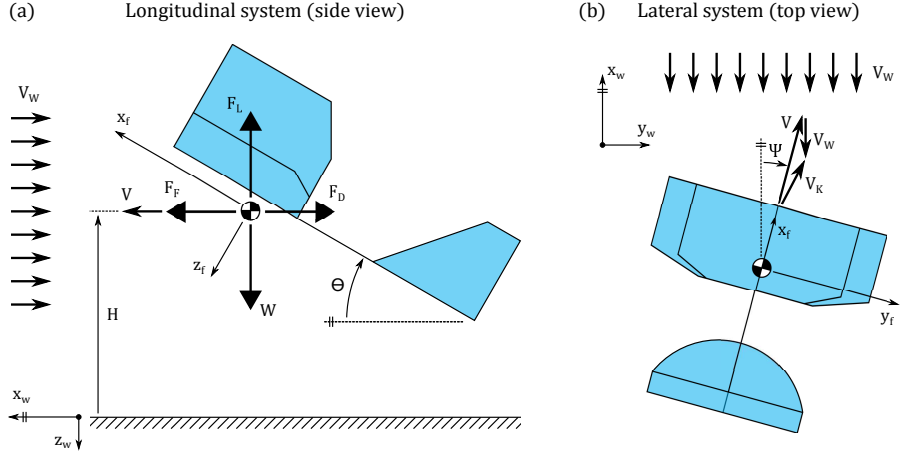}
\caption{Definition of the axis systems. Two frames, wind-tunnel-fixed $w$ and body-fixed $f$, are introduced to define the body position in the wind-tunnel and the body attitude angles, respectively. Consistent with the aerospace convention, the z axis is pointing downwards. Panel (a) displays the side view with the longitudinal system parameters, assuming steady flight against the free stream $V_W$. Panel (b) shows the top view with the lateral system parameters. Due to no roll control authority, displacement in the $y_w$ direction is achieved through heading $\Psi$.}
\label{fig:axis_system}
\end{figure}

The body position $\mathbf{x}$ is expressed in the ground fixed system aligned with the wind tunnel: the $x_w$-axis points opposite the wind velocity vector, $z_w$ points down and $y_w$ completes the right-handed Cartesian system, see Fig.~\ref{fig:axis_system}. The body-fixed coordinate system is defined by the body's main axes: the $x_f$-axis points along the fuselage towards the nose, the $z_f$-axis points opposite to the direction of the vertical stabiliser, and the $y_f$-axis points starboard. Its origin is placed at the centre of gravity. Because the external motion tracking system measures the position of the geometrical centre of the four LED markers, we used that value as an approximation of the centre of mass position. The body attitude $\mathbf{\Phi}$ is described by roll $\Phi$, pitch $\Theta$ and yaw $\Psi$ angles, which define the rotation around the $x_f$, $y_f$ and $z_f$ axes, respectively.

The aircraft velocity $\vec{V}$ is (in steady-state) pointing opposite to the wind velocity vector $\vec{V}_W$, that is we have $\vec{V} = -\vec{V}_W$ and there is no motion relative to ground ($\vec{V}_K = \vec{V} + \vec{V}_W = \vec{0}$, see the longitudinal system in figure~\ref{fig:axis_system} a)). Height $H = -z_w$ is used as a measure of vertical position. The lift force $F_L$ and thrust force $F_F$ are oriented normal and parallel to the wind velocity $V_W$, respectively. They act at the centre of gravity and, in steady state, compensate the weight $W$ and drag force $F_D$.

Figure~\ref{fig:axis_system} b) shows the lateral system for the case of non-zero heading $\Psi$. In such case the aircraft will move relative to ground with a non-zero velocity $\vec{V}_K = \vec{V} + \vec{V}_W$. 

\subsubsection{Control overview}

\begin{figure}[h!]
\centering
\includegraphics[width=\linewidth]{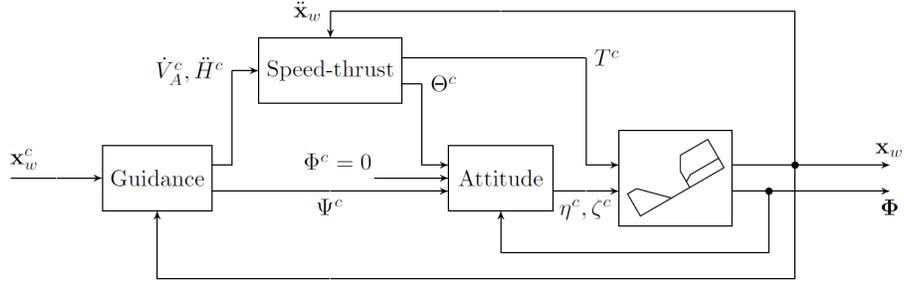}
\caption{Block diagram of the cascade control approach consisting of guidance, speed-thrust and attitude controllers. Guidance block commands the desired heading $\Psi$ and accelerations $\dot{V}_A^c, \ddot{H}^c$ based on the current position error. The speed-thrust block determines the combination of pitch $\Theta^c$ and thrust $T^c$ commands that leads, at the wind tunnel speed $V_W$, to the desired accelerations. The attitude block controls the attitude through the rudder and elevator commands, $\eta^c$ and $\zeta^c$, respectively.}
\label{fig:control_approach}
\end{figure}

In the wind-tunnel experiment, the desired flight path is simply a (ground-) fixed way-point. Since the flight dynamics of the DelFly are still being investigated and the linearised models identified so far are only valid at a single operating condition \cite{Armanini2016}, no reliable model that would cover the whole flight envelope was available. Thus, we employed a traditional aerospace control approach with control loops in a cascade arrangement, as implemented in the open-source Paparazzi UAV System \cite{Paparazzi}. However, an additional speed-thrust control block was added in between the standard guidance and attitude control blocks to take care of varying thrust and lift produced at different body speeds (and body attitudes), see figure \ref{fig:control_approach}. Thus, the guidance control determines the desired body accelerations and heading based on the position error from the set-point. The commanded accelerations are transformed into the desired thrust and pitch by the speed-thrust control block. While novel to FWMAVs, a similar solution was used in the transitioning phase of hybrid UAVs \cite{Theys2016,DeWagter2013}. Finally, the attitude control loop determines the rudder and elevator deflections necessary to achieve the desired body pitch and heading.

\subsubsection{Guidance control}

The guidance control is decentralised, i.e. we control the forward position $x_w$, height $H = -z_w$ and lateral position $y_w$ in separate loops. In the longitudinal loops (forward + vertical) we assume the DelFly is always aligned with the wind tunnel axis, i.e. it is flying opposite to the wind direction $\vec{V}_W$. Ordinary PD controllers are used in the longitudinal and vertical loops to determine the desired accelerations $\dot{V}^c$ and $\ddot{H}^c$, which are commanded to the inner loops. Since the DelFly has no roll control authority, the lateral position $y_w$ is controlled through heading $\Psi^c$. To compensate for steady state errors, an integral gain was introduced to the lateral loop. 

The accelerations and heading commanded to the inner loops are thus determined as
\begin{align}
    \dot{V}^c &= k_{dx} \Delta\dot{x}_w + k_{px} \Delta x_w \\
    \ddot{H}^c &= - k_{dz} \Delta\dot{z}_w - k_{pz} \Delta z_w \\
    \Psi^c & = k_{dy} \Delta\dot{y}_w + k_{py} \Delta y_w + k_{iy} \int \Delta y_w \dd{t},
\end{align}
where $k_p$, $k_d$ and $k_i$ denotes the P, D and I gains, respectively, and $\Delta$ stands for the position error from the set-point.

The P and D gains of the longitudinal and vertical loops were selected based on the desired closed loop behaviour (assuming the plant to be a second order system with no inherent damping). The gains of the lateral loop were tuned during the flight tests. All the gain values, as used for the experiments in section \ref{sec:results_position}, are summarised in table \ref{tab:gain_values}.

\subsubsection{Speed-thrust control}
\label{sec:speed_thrust}

\begin{figure}[p]
	\centering
    \includegraphics[width=\linewidth]{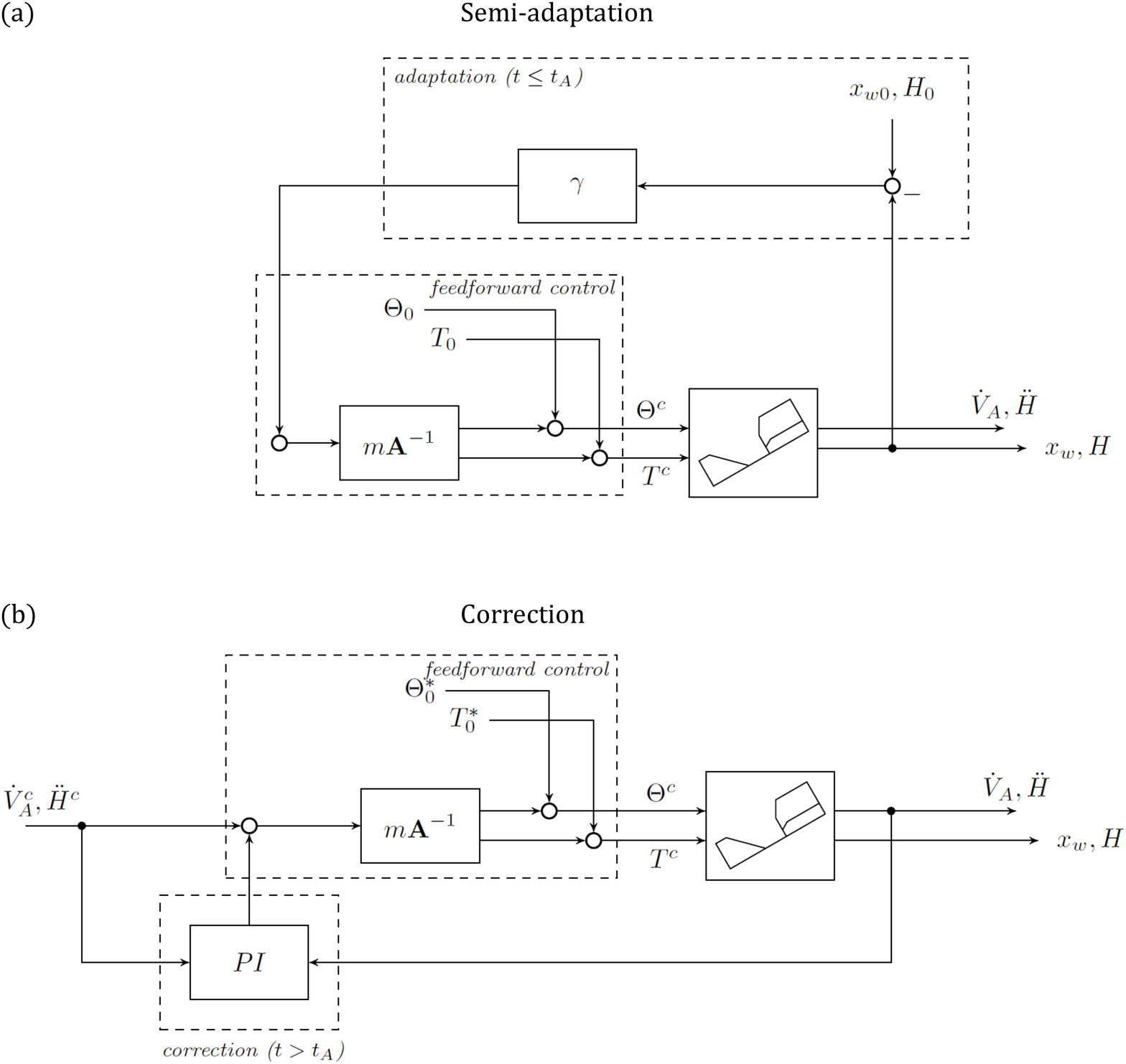}	
    \caption{The two-phase semi-adaptive control approach. At the start of each flight, an adaptation loop with gain $\gamma$ is used to adapt the assumed equilibrium conditions $T_0$ and $\Theta_0$, until any potential position drift caused by model uncertainties of the feedforward control is removed. At time $t = t_A$, when the equilibrium is approached, the operator switches to the correction phase and the adapted equilibrium conditions $T_0^*$ and $\Theta_0^*$ are kept. In this phase the acceleration set-points from the outer guidance loop are tracked and an additional feedback loop is employed to compensate for disturbances, model uncertainties and performance changes due to decreasing battery voltage.}
	\label{fig:semiadaptive}
\end{figure}

Since the generation of lift and thrust is highly coupled, a suitable combination of pitch angle $\Theta$ and throttle command $T$ (controlling the flapping frequency $f$) that will result in the desired accelerations in the longitudinal and vertical directions needs to be found. This is the role of the speed-thrust control (figure \ref{fig:semiadaptive}), which consists of a feedforward and feed-back part. The feedforward control selects the necessary combination of $\Theta$ and $T$ based on a linear model constructed from wind tunnel force measurements data. The feedback part improves the performance by correcting for model uncertainties, change of performance over time as well as external disturbances.

\paragraph{Feedforward control}
Using wind tunnel data obtained with a clamped DelFly for various wind speeds $V_W$, pitch angles $\Theta$, and throttle commands $T$ (the data were collected during an experiment described in \cite{Karasek2016}), a linear relationship between the pitch angle and throttle and the measured thrust and lift forces can be found by first-order Taylor linearisation
\begin{align}
	\label{eq:speedthrust-liftanddrag}
	\begin{bmatrix} F_F \\ F_L \end{bmatrix} = \begin{bmatrix} F_{F0}(V_W) \\ F_{L0}(V_W) \end{bmatrix} + A(V_W) \begin{bmatrix} \Theta - \Theta_0(V_W) \\ T - T_0(V_W) \end{bmatrix},
\end{align}
where $\Theta_0(V_W), T_0(V_W)$ denote the equilibrium condition of no acceleration for wind speed $V_W$, resulting into a thrust $F_{F0}(V_W)$ that is equal to drag at $V_W$ and lift $F_{L0}(V_W)$ that is equal to the DelFly weight. $A(V_W)$ is the matrix of force derivatives (evaluated at the equilibrium $\Theta_0(V_W), T_0(V_W)$)
\begin{equation}
    A(V_W) = \begin{bmatrix} \frac{\partial F_F}{\partial \Theta}(V_W) & \frac{\partial F_F}{\partial T}(V_W) \\ \frac{\partial F_L}{\partial \Theta}(V_W) & \frac{\partial F_F}{\partial T}(V_W) \end{bmatrix}.
\end{equation}

\noindent 
By inverting equation~(\ref{eq:speedthrust-liftanddrag}) we get
\begin{align}
	\begin{bmatrix} \Theta \\ T \end{bmatrix} = \begin{bmatrix} \Theta_0(V_W) \\ T_0(V_W) \end{bmatrix} + A(V_W)^{-1} \begin{bmatrix} \Delta F_F(V_W) \\ \Delta F_L(V_W) \end{bmatrix},
\end{align}
where $\Delta F_F(V_W)$ and $\Delta F_L(V_W)$ are the differences of thrust and lift from the equilibrium values $F_{F0}(V_W)$ and $F_{L0}(V_W)$, respectively. Assuming the horizontal and vertical systems are decoupled and neglecting any inherent damping, the body acceleration is a result of only the forces applied, $\dot{V}_A=\Delta F_F/m, \ddot{H}=\Delta F_L/m$. Thus, the desired pitch angle and thrust can be found from the acceleration set-points as
\begin{align}
	\begin{bmatrix} \Theta^c \\ T^c \end{bmatrix} = \begin{bmatrix} \Theta_0(V_W) \\ T_0(V_W) \end{bmatrix} + m A(V_W)^{-1} \begin{bmatrix} \dot{V}^{sp} \\ \ddot{H}^{sp} \end{bmatrix}.
\end{align}

The matrix of aerodynamic force derivatives has been derived for different wind speeds, see table \ref{tab:aero_derivatives}; switching between the different values is done manually based on the wind tunnel set-point, which remains constant throughout the tests.

\begin{table}[h]
\centering
\caption{Equilibrium conditions and aerodynamic force derivatives for various wind speeds, based on wind tunnel measurements described in \cite{Karasek2016}.}
\label{tab:aero_derivatives}
\begin{tabular}{ c||c|c||c|c|c|c }
    $V_W$ &  $\Theta_0$ & $T_0$ & $\frac{\partial F_F}{\partial \Theta}$ & $\frac{\partial F_F}{\partial T}$ & $\frac{\partial F_L}{\partial \Theta}$ & $\frac{\partial F_L}{\partial T}$ \\
    (m/s) &  ($^\circ$) & (\%) & (mN/$^\circ$) & (mN/\%) & (mN/$^\circ$) & (mN/\%) \\
    \hline
    0.8 &  65.9 & 86.8 & -5.2 & 1.4 & 0.8 & 3.7 \\
    1.2 &  47.2 & 78.0 & -2.8 & 2.4 & 0.8 & 3.4 \\
    2.5 &  30.5 & 68.5 & -5.5 & 2.2 & 4.9 & 3.2 \\
\end{tabular}
\end{table}

\paragraph{Semi-adaptation}
In ideal case, setting the throttle and elevator to the equilibrium values $\Theta_0(V_W), T_0(V_W)$ should result in steady state flight at speed $V_W$. However, because the wind tunnel data used to derive the matrices of force derivatives $A(V_W)$ were obtained with another DelFly, and because the performance of the DelFly can deteriorate over time, the DelFly will typically drift (with respect to ground). To correct for the drift, each flight started with an adaptation phase, during which the assumed equilibrium conditions $\Theta_0(V_W), T_0(V_W)$ are being adapted, via a position feedback, until a steady flight is reached
\begin{align}
	\begin{bmatrix} \Theta_0^*(V_W) \\ T_0^*(V_W) \end{bmatrix} = \begin{bmatrix} \Theta_0(V_W) \\ T_0(V_W) \end{bmatrix} + m A(V_W)^{-1}\gamma \begin{bmatrix} \Delta x_w \\ \Delta H \end{bmatrix},
\end{align}
where the $^*$ superscript denotes the adapted equilibrium conditions, $\Delta x_w$ and $\Delta H$ are errors from the position where the adaptation was started and $\gamma$ is a positive gain. The proof that such feedback leads to a stable equilibrium is in the Appendix of \cite{Cunis2016}.

\paragraph{Feedback control}
Once the equilibrium is found, the operator switches to a correction phase, where an additional feedback loop with a PI controller is added to the feedforward control to compensate for disturbances, model uncertainties and performance changes due to decreasing battery level

\begin{align}
	\begin{bmatrix} \dot{V}^\text{fb} \\ \ddot{H}^\text{fb} \end{bmatrix} = \begin{bmatrix} k_{pF} \Delta \dot{V} \\ k_{pV} \Delta \ddot{H}\end{bmatrix} + \begin{bmatrix} k_{iF} \int \Delta \dot{V} \dd t \\ k_{iV} \int \Delta \ddot{H} \dd t \end{bmatrix}.
\end{align}
The integrated error is calculated as $\int\Delta\dot{V}\dd t = \int{\dot{V}^{sp}}\dd t - V_A, \int\Delta\ddot{H}\dd t = \int{\ddot{H}^{sp}}\dd t-\dot{H}$. 

Because $A$ contains a guess of direction of the force derivatives, we add the correction before the feedforward control. The combined control law results into
\begin{align}
	\begin{bmatrix} \Theta^c \\ T^c \end{bmatrix} = \begin{bmatrix} \Theta_0^*(V_W) \\ T_0^*(V_W) \end{bmatrix} + m A^{-1}(V_W) \left ( \begin{bmatrix}\dot{V}^{sp} \\ \ddot{H}^{sp} \end{bmatrix} + \begin{bmatrix} k_{pF} \Delta \dot{V} \\ k_{pV} \Delta \ddot{H} \end{bmatrix} + \begin{bmatrix} \int k_{iF} \Delta\dot{V} \dd{t} \\ \int k_{iV} \Delta\ddot{H} \dd{t} \end{bmatrix} \right ).
\end{align}

\subsubsection{Attitude control}

In the inner most loop, we used the Integer-quaternion implementation of the attitude stabilisation algorithm of the Paparazzi UAV system \cite{Paparazzi}, which controlled the body pitch and yaw via elevator and rudder deflections, respectively. The PID gains were tuned manually prior to the wind tunnel tests. For faster speeds a feed-forward term $k_{ff}$ was used in the yaw loop. The gain values are summarised in table \ref{tab:gain_values}.

\begin{table}[h]
\centering
\caption{Wind speed dependent gain values of all the control loops. The high speed gains are significantly different because the FWMAV gets close to the limit of inherent stability at these speeds.}
\label{tab:gain_values}
\begin{tabular}{ c || c | c || c }
     & \multicolumn{2}{c||}{Low speed} & High speed \\
    $V_W$ &  $\approx$ 0.8 m/s & $\approx$ 1.2 m/s & $\approx$ 2 to 2.4 m/s \\
    \hline
    & \multicolumn{3}{c}{Guidance control} \\
    \hline
    $k_{px}$ & \multicolumn{3}{c}{1} \\
    $k_{dx}$ & \multicolumn{3}{c}{2} \\
    $k_{py}$ & 6 & 4.8 & 6 \\
    $k_{iy}$ & 0.75 & 0.6 & 0.98 \\
    $k_{dy}$ & 0.75 & 1.2 & 3 \\
    $k_{pz}$ & \multicolumn{3}{c}{1} \\
    $k_{dz}$ & \multicolumn{3}{c}{2} \\
    \hline
    & \multicolumn{3}{c}{Speed-thrust control} \\
    \hline
    $k_{pv}$ & \multicolumn{2}{c||}{0} & 0.3  \\
    $k_{iv}$ & \multicolumn{2}{c||}{2} & 0.5 \\
    $k_{pf}$ & \multicolumn{2}{c||}{0} & 1 \\
    $k_{if}$ & \multicolumn{2}{c||}{2} & 0.3 \\
    $\gamma$ & \multicolumn{3}{c}{2.5} \\
    \hline
    & \multicolumn{3}{c}{Attitude control} \\
    \hline
    $k_{p\Theta}$ & \multicolumn{2}{c||}{2.53} & 1.75 \\
    $k_{i\Theta}$ & \multicolumn{2}{c||}{0.25} & 0.034 \\
    $k_{d\Theta}$ & \multicolumn{2}{c||}{0.063} & 0.031 \\
    $k_{p\Psi}$ & \multicolumn{2}{c||}{1.63} & 1.5 \\
    $k_{i\Psi}$ & \multicolumn{2}{c||}{0.019} & 0 \\
    $k_{d\Psi}$ & \multicolumn{2}{c||}{0.094} & 0.031 \\
    $k_{ff\Psi}$ & \multicolumn{2}{c||}{0} & 5 \\
\end{tabular}
\end{table}

\subsection{PIV measurement setup and processing}
\label{sec:methods_PIV}

High-speed Stereo-PIV measurements were performed at a spanwise-oriented plane approximately 150 mm downstream from the DelFly tail. Note that only one side of the wake was imaged, due to field of view size restrictions, however, the wake is assumed to be nominally symmetric with respect to the centre plane. The flow was illuminated with a double-pulse Nd:YLF laser (Mesa-PIV) with a wavelength of 527 nm. The laser sheet with a thickness of 2 mm was kept at a fixed position, while the DelFly position was varied based on the measurement case. The flow was seeded with a water-glycol based fog of droplets with a mean diameter of 1 $\upmu$m, which is produced by a SAFEX fog generator. The complete measurement room was filled with the fog beforehand in order to achieve a homogeneous seeding of the flow. Images of tracer particles were captured with two high-speed Photron FastCam CMOS cameras which allows to achieve a maximum resolution of 1024 $\times$ 1024 pixels at a data rate of up to 5.4 kHz. Each camera was equipped with a Nikon 60 mm focal objective with numerical aperture f/4 and mounted with Scheimpflug adaptors. The cameras were placed with an angle of 40$^\circ$ with respect to each other. A schematic overview of the PIV measurement setup can be seen in figure \ref{fig:experimental_setup}. 

A field of view of 170 mm $\times$ 170 mm was captured with a magnification factor of approximately 0.12 at a digital resolution of 6 pixels/mm. Single-frame images were recorded at a rate of 5 kHz for approximately a second, yielding a data ensemble size of about 5000 images. The associated time interval of 0.2 ms between individual frames corresponds to an out-of-plane displacement of 0.4 mm, based on the free stream velocity. This is sufficiently low with respect to the laser sheet thickness to allow for an accurate correlation of subsequent particle images. In order to increase the signal-to-noise ratio in the images, two laser cavities were shot in each single camera frame with 1 $\upmu$s time delay in between, ensuring frozen particle images. The commercial software Davis 8.0 (LaVision) was used in data acquisition, image pre-processing, stereoscopic correlation of the images, and further vector post-processing. The pre-processed single-frame images were interrogated using windows of final size of 64 $\times$ 64 pixels with two refinement steps and an overlap factor of 75$\%$ resulting in approximately 4800 vectors with a spacing of 2.9 mm in each direction.

A spatio-temporal reconstruction was performed for the initial interpretation of the wake structure. For this purpose, the time-series measurements performed in the single static measurement plane (i.c., around 150 mm downstream of the DelFly tail) were employed to generate a quasi-three-dimensional representation of the wake structure by using a passive convection model (Taylor’s hypothesis). This implies that the data of the measurement plane is translated with the free-stream velocity $U_{\infty}=V_W$ (with an assumption of non-deforming wake and neglecting the induced velocities). More precisely, from the time-resolved velocity data $u(x,y,z,t)$ recorded at the fixed streamwise position $x=x_0$, the volumetric representation of the instantaneous flow field at a specific time $t=t_0$ is computed as $u(x,y,z,t_0) \approx u(x_0,y,z,t_0-x/U_{\infty})$.
This approach results in a spatial resolution of 0.4 - 0.5 mm in the streamwise direction for the given image recording rate of 5 kHz and for the considered free-stream velocity range (2 - 2.4 m/s).

\section{Results}
\label{sec:results}

\subsection{Position control}
\label{sec:results_position}

The following section shows the performance of the position control, in steady state as well as in response to a step input. During the PIV measurement, the flying DelFly should ideally stay at a prescribed constant position. Thus, our primary goal was to achieve high precision steady flight around a fixed set-point rather than fast tracking performance of a moving set-point.

\subsubsection{Step commands}
\begin{figure}[p]
	\centering
    \includegraphics[width=0.7\linewidth]{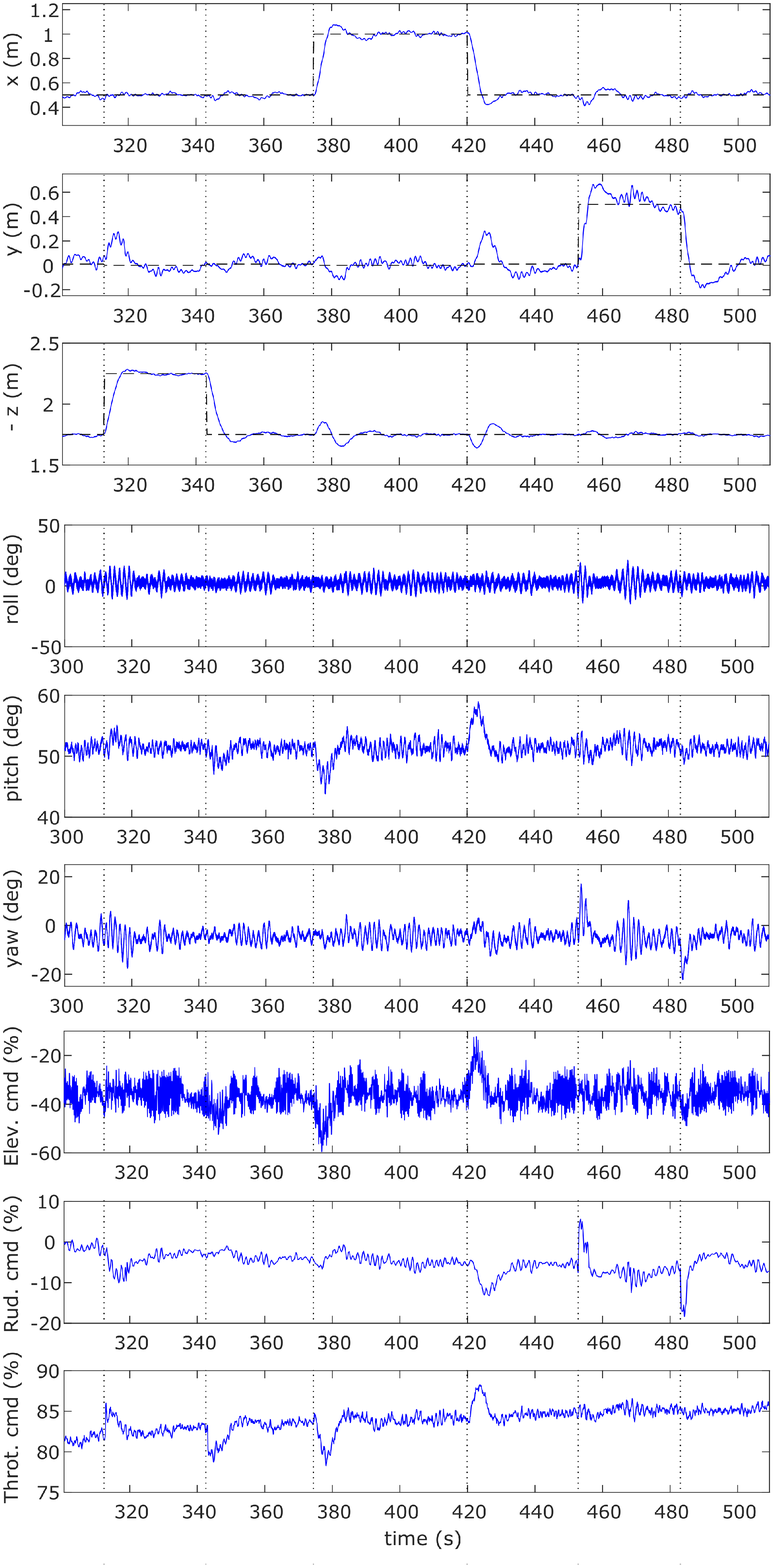}	
    \caption{Response to a sequence of step commands in all three directions at $V_W$ = 1.2 m/s. The position set-points are displayed in black dashed lines, blue lines show the unfiltered tracking and telemetry data. The dotted vertical lines mark the time stamps of the step commands.}
	\label{fig:step}
\end{figure}

The sequence of step commands in all three wind tunnel frame axes is captured, for a wind tunnel set-point $V_W$ = 1.2 m/s, in figure \ref{fig:step}. Apart from the position, we also recorded the body attitude (motion tracking system) and commands to the attitude loop and to the motor speed controller (WiFi telemetry). From the position graphs we can see that similar rise times, between 3 and 6 seconds, were achieved in all the three directions. A slight overshoot and longer settling was observed especially in the lateral direction, since it was controlled indirectly, through the change of heading.

In longitudinal manoeuvres, it is the speed thrust control block (section \ref{sec:speed_thrust}) that determines the necessary combination of throttle and elevator commands, based on current wind tunnel set-point. In figure \ref{fig:step} ($V_W$ = 1.2 m/s), the vertical manoeuvre is dominated by a throttle change, while forward manoeuvre also requires pitching the body via the elevator. Because this block is based on experimental data obtained with a slightly different aircraft, some coupling remains when forward step is commanded, nevertheless the feedback control damps these effects out. The lateral position is sensitive to both changes in vertical and forward directions, which is an inherent property of the aircraft, but again the feedback control will steer the vehicle back to the set-point through a heading change controlled by rudder deflection. 

From the command plots we can further observe that the throttle command increases over time. This is due to the battery voltage, which is decreasing as the battery gets discharged, and due to the integrator action, which responds by increasing the throttle command in order to keep the flapping frequency constant. A comparison of measured body pitch with the pitch command confirms that the attitude loop manages to follow the pitch set-point. A post flight telemetry analysis showed that the decreasing trend in the yaw command is a result of a slow drift of the on-board heading estimate. While heading from the tracking system should be used for correcting the drift typical when integrating gyroscope readings, a small drift remained and was again corrected for by the integrators in the control loops. The heading measured by the tracking system remained close to zero, i.e. aligned with the wind tunnel axis.

\subsubsection{Steady state}
\begin{figure}[p]
	\centering
    \includegraphics[width=\linewidth]{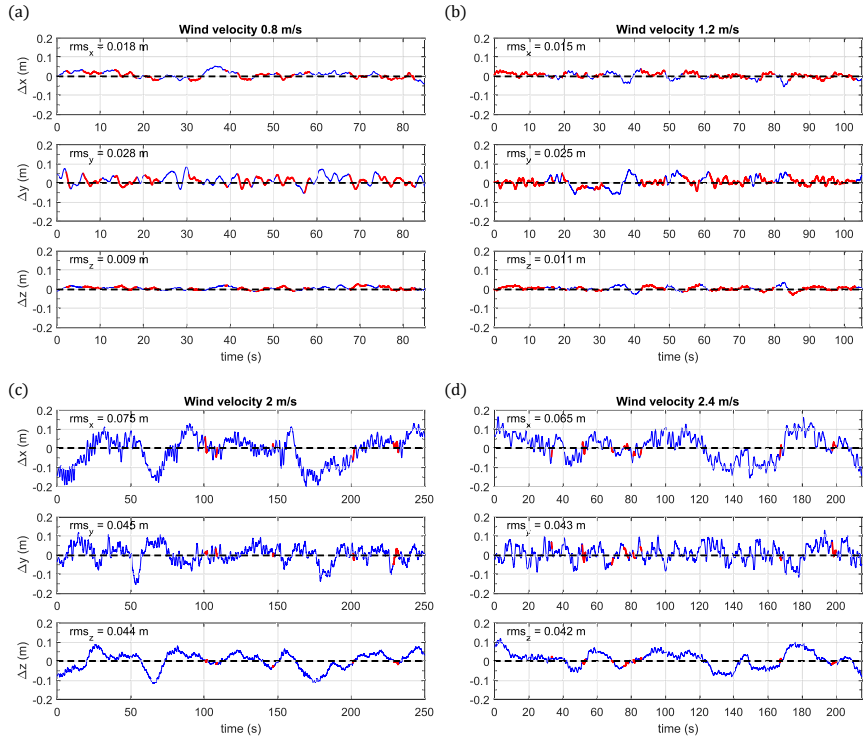}
    \caption{Position difference from the set-point for various wind tunnel speeds: (a) 0.8 m/s, (b) 1.2 m/s, (c) 2 m/s, (d) 2.4 m/s. Root-mean-square ($rms$) position error values over the whole measurement are displayed in the top left corner of each subplot. The segments where a successful PIV measurement could be performed according to our estimations (i.e. the $rms$ error remains below 25 mm in all three axes for the next 2 seconds) are highlighted by a thicker red line. Significant position accuracy decrease can be observed for high speeds (2 m/s and 2.4 m/s), where the FWMAV is operated close to its inherent stability limit. Besides, the gain tuning for high speeds may not have been optimal due to time constraints during the wind-tunnel slot dedicated to the flow visualisation measurements.}
	\label{fig:steady_state}
\end{figure}

The results of steady flight with a fixed position set-point, performed at wind tunnel set-points $V_W$ = 0.8 m/s, 1.2 m/s, 2.0 m/s and 2.4 m/s, are in figure \ref{fig:steady_state}. Each panel shows the difference from the set-point in all three wind tunnel axes. The highlighted parts (thicker line, red colour) show the segments, where a successful PIV measurement could be performed according to our estimations, i.e. where the root mean square error ($rms$) remains below 25 mm in all three axes for the next 2 seconds. The $rms$ errors over the whole measurement are summarised, together with mean body attitude angles, mean marker tracking errors and their respective standard deviations, in table \ref{tab:steady_state}. All the data was measured by the motion tracking system.

\begin{table}[h]
\centering
\caption{Position precision and attitude at various wind tunnel set-points. The attitude and mean marker tracking error are represented as mean $\pm$ standard deviation over the interval displayed in figure \ref{fig:steady_state}.}
\label{tab:steady_state}
\begin{tabular}{ c || c | c | c || c | c | c || c}
    $V_W$ &  $rms_x$ & $rms_y$ & $rms_z$ &  $\Phi$ & $\Theta$ & $\Psi$ & Track. err. \\
    (m/s) &  (mm) & (mm) & (mm) &  ($^\circ$) & ($^\circ$) & ($^\circ$) & (mm)\\
    \hline
    0.8 & 18 & 28 & 9 & 4.5 $\pm$ 3.4 & 68.9 $\pm$ 0.8 & 0.3 $\pm$ 3.4 & 0.73 $\pm$ 0.17 \\ 
    1.2 & 15 & 25 & 11 & 2.3 $\pm$ 3.1 & 50.5 $\pm$ 0.8 & -3.1 $\pm$ 1.7 & 0.45 $\pm$ 0.11 \\
    2.0 & 75 & 45 & 44 & 1.5 $\pm$ 2.7 & 33.8 $\pm$ 1.4 & -3.4 $\pm$ 1.5 & 0.76 $\pm$ 0.30 \\
    2.4 & 65 & 43 & 42 & 1.7 $\pm$ 2.9 & 28.7 $\pm$ 1.1 & -3.6 $\pm$ 1.4 & 0.59 $\pm$ 0.19 \\
\end{tabular}
\end{table}

It can be immediately noted that the performance at low speeds is much better than at high speeds. Originally, prior to the PIV test session, we tuned the controller for speeds ranging from 0.8 m/s to 1.2 m/s, where the inherent stability of the DelFly is the most pronounced. However, during the PIV session we observed that the quality of captured data was worse than expected. The direction of the wake structures was dominated by the flapping-induced flow (aligned with the body fuselage that is pitched by 50.5$^\circ$ to 68.6$^\circ$ at slow speeds) and this resulted into a considerable angle between the measurement plane normal and the wake axis. Therefore, within the time constraints of the wind tunnel slot available for the PIV tests, we quickly tuned the controller also for high speeds (2.0 m/s and 2.4 m/s shown here), where the body pitch is much lower (33.8$^\circ$ to 28.7$^\circ$). This was much more challenging, because the DelFly (in the configuration used for the tests) is already very hard to fly at these speeds without any stability augmentation. 

At low speeds (0.8 m/s and 1.2 m/s), the position fix was very good. The best results were achieved in the vertical $z_w$ direction, where the DelFly stayed within $\pm$ 25 mm most of the time and the corresponding $rms$ error was around 10 mm. In the forward direction, an accuracy better than $\pm$ 50 mm was achieved most of the time, with an $rms$ error of below 20 mm. Most oscillations were observed in the lateral direction, yet a large part of the time the aircraft was also within $\pm$ 50 mm from the set-point, which can be seen from the rms values that remain below 30 mm. According to the estimated criteria ($rms$ below 25~mm for the next 2 seconds) a successful PIV measurement could be started 49\% and 65\% of the time for 0.8 m/s and 1.2 m/s, respectively, meaning the waiting time of the operator monitoring the MAV position and triggering the PIV measurement would be very short.

At high speeds, the DelFly control is much more challenging as explained earlier. The $rms$ error values were about 45 mm in forward and lateral directions and around 70 mm for forward direction. Despite a worse overall performance (the rms values were computed over several minutes of flight), there were segments of several seconds (highlighted parts) when the platform was very close to the set-point for at least 2 seconds in all the axes (3\% and 4.8\% of the total time for 2.0~m/s and 2.4~m/s, respectively). This gave us enough opportunities to trigger and perform successful PIV measurements also at lower body pitch angles, where the flow patterns of the wake move almost normally to the vertical measurement plane, yielding higher quality flow measurements.

The mean attitude captured at different speeds (table \ref{tab:steady_state}) reveals that the roll and yaw angles were not exactly zero. This is due to imperfections of the hand-built DelFly, in particular a slight misalignment of the tail caused by a twist in the square carbon tube used as fuselage, but this has a negligible effect on the PIV measurements since the misalignment is in the order of a few degrees.

\subsection{Flow visualisation}

This section provides results for the first free-flight flow visualization of the wake of the DelFly. The PIV measurements were performed in a plane oriented perpendicular to the freestream direction, at a distance of approximately 150 mm downstream of the tail, similar to measurements performed with bats of comparable sizes \cite{Hedenstrom2009}. Results are presented here for the flight condition of a freestream speed of 2.4 m/s, flapping frequency of 12.0 Hz and body angle of 28.7$^\circ$. The corresponding reduced frequency, defined as $k = \pi fc/V_W$ (where $f$ is the flapping frequency, $c$ = 80~mm the mean wing chord and $V_W$ the freestream velocity) has a value of 1.25. The Reynolds number is 13,000, based on freestream velocity and wing chord. For comparison, the tests were also conducted in a tethered setting with the identical DelFly, rigidly fixed according to figure \ref{fig:setup_photo}, similar to our previous trials \cite{Percin2014}. The flight conditions were comparable to the free-flight tests (freestream speed of 2.0 m/s, flapping frequency of 12.1 Hz, body angle of 33.8$^\circ$ and Reynolds number around 11,000).

PIV images were recorded for a duration of approximately one second, at an acquisition frequency of 5 kHz. Given the flapping frequency of 12 Hz, this implies that 12 cycles are captured, with approximately 400 images per cycle, indicating a well-resolved characterisation of the flapping cycle. The results presented in the following are the direct outcome of the measurements, i.e., no additional averaging, smoothing or other form of filtering has been applied that could potentially further improve the quality of the visualisation.

\begin{figure}[h]
	\centering
    \includegraphics[width=\linewidth]{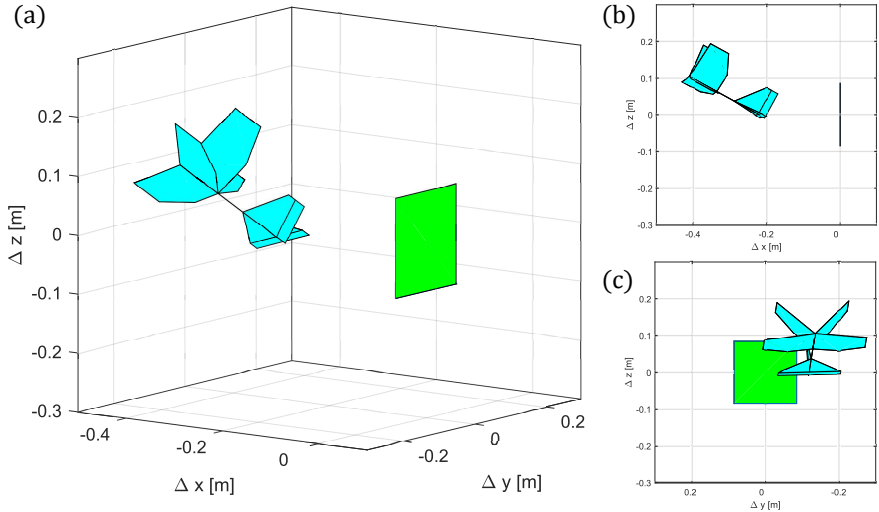}
    \caption{Relative position and orientation of the FWMAV with respect to the measurement plane (green square), averaged over the duration of the PIV measurement presented in figures \ref{fig:PIV_VorticityPlanes} and \ref{fig:PIV_Helicity}.}
	\label{fig:PIV_position}
\end{figure}

The relative position and orientation of the FWMAV with respect to the centre of the measurement region, averaged over the duration of the PIV recording, is displayed in figure \ref{fig:PIV_position}. While various position set-points were tested during the trials, this relative position allowed to capture the most prominent vortex structures in the wake, originating from the right halve of the wings. The good position stability over the duration of the PIV measurement can be documented by the values of standard deviation from the mean position, which show that apart from a slight drift in the y direction ($\sigma_y = 26.3$ mm) a very good fix was achieved both in the forward ($\sigma_x = 5.2$ mm) and the vertical ($\sigma_z = 4.5$ mm) axes.

\begin{figure}[p]
\centering
\includegraphics[width=0.8\linewidth]{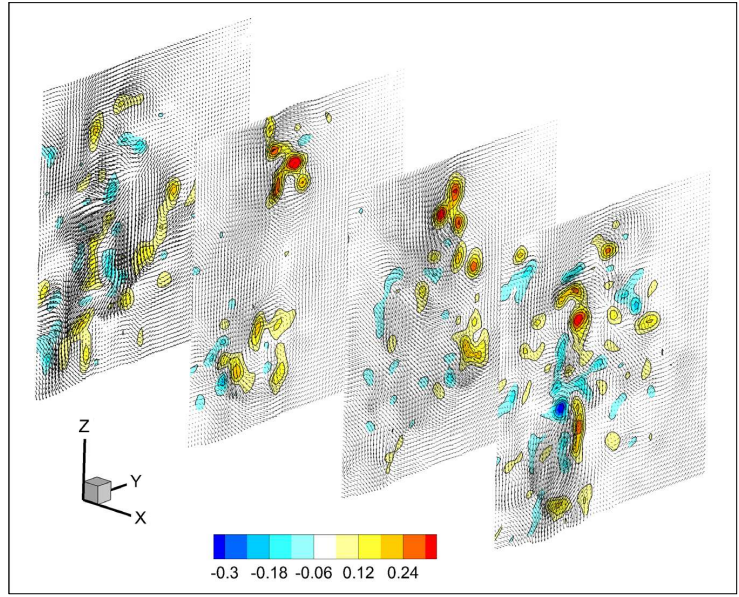}
\caption{Stereo-PIV measurements in the wake of a DelFly in free flight showing a sample sequence of four images, separated by 0.065 seconds, where the rightmost one is captured earliest in time; vectors indicate the in-plane velocity components and the colour contours the out-of-plane vorticity (in 1/s). Free stream velocity 2.4 m/s; flapping frequency is 12.0 Hz.}
\label{fig:PIV_VorticityPlanes}
\end{figure}

Figure \ref{fig:PIV_VorticityPlanes} displays a sample time series of four images separated by 0.065 seconds, with the vectors indicating the in-plane velocity components and the colour contours the out-of-plane vorticity. The most prominent feature observed in the visualisations can be associated to the tip vortex of the upper wing in the instroke phase (red, corresponding to counter-clockwise vorticity).

\begin{figure}[p]
\centering
\includegraphics[width=\linewidth]{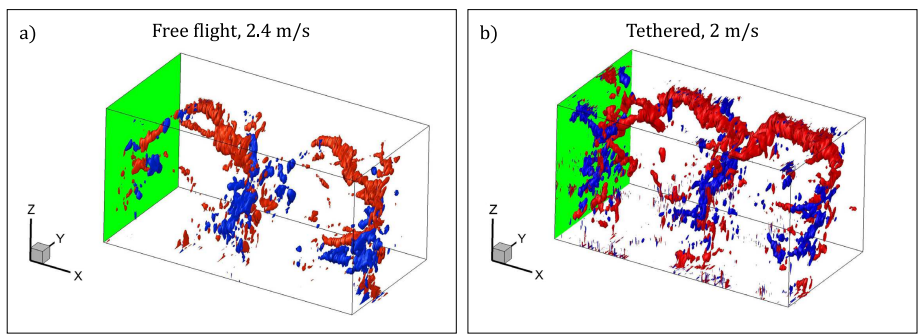}
\caption{Comparison of three-dimensional wake structure of the DelFly for (a) free-flight and (b) tethered condition; colour coding is for helicity (red: +0.6 m/s$^2$; blue: -0.6 m/s$^2$).}
\label{fig:PIV_Helicity}
\end{figure}

A three-dimensional representation of the wake vortex structure is obtained with the convection model, as described in section \ref{sec:methods_PIV}, which transforms the temporal information contained in the high-speed velocity field acquisition into a spatial representation by translating the flow field of subsequent images downstream with the freestream velocity. The result, displaying two flapping cycles, is shown in figure \ref{fig:PIV_Helicity}. Figure \ref{fig:PIV_Helicity}a applies to the free-flight condition and figure \ref{fig:PIV_Helicity}b to the tethered DelFly. The visualisations provide a colour-coding of the helicity (density), which is defined as the scalar product of the velocity and vorticity vectors. 

Helicity can be used for the detection of vortex cores \cite{Degani1990} and non-zero helicity indicates a helical vortex structure with an axial flow. The sign of the helicity allows to distinguish vortex structures with different sense of swirl. The very prominent upper red structure (positive helicity) is the tip vortex formed by the upper wing during the instroke. The less distinct blue structure (negative helicity) is the tip vortex of the bottom wing generated during instroke as well. Structures of the outstroke appear not to be very well captured in this representation, however.

Notwithstanding the suboptimal quality of these preliminary results, an important observation is the good qualitative agreement between the free-flight and tethered wake flow structures, and the good repeatability of the two cycles for each case. This supports the conclusion that reliable and meaningful PIV measurement results have been obtained also in the free-flight case. Further processing of the data may allow for a more quantitative and detailed comparison between the free flight and tethered condition and potentially reveal if there are effects of the tethering to be detected. 


\section{Conclusions and discussion}
\label{sec:conclusions}
We presented a methodology, which combined a FWMAV specific control approach for autonomous flight in a wind tunnel with a time-resolved stereoscopic PIV and allowed the first flow visualisation experiments to be carried out with a freely flying flapping-wing robot. The novel FWMAV specific control approach relied on feedback from an on-board IMU and an external motion tracking system. Applied to the 23 g DelFly FWMAV, an autonomous flight with high accuracy at low speeds (0.8 - 1.2 m/s, maximal root-mean-square error of 28 mm over 1-2 minutes) and good accuracy at high speeds (2 - 2.4 m/s, maximal root-mean-square error of 75 mm over 3-4 minutes) was achieved. Moreover, even higher precision was often achieved for time intervals of several seconds. Thus, the PIV measurements, lasting around one second, could be triggered when the DelFly was at the ideal position, which permitted to use a smaller measurement region and resulted in high resolution flow data.

The free-flight PIV measurements were performed at high free stream speeds (2 to 2.4 m/s), where the FWMAV is pitched by 33.8$^\circ$ to 28.7$^\circ$ and flaps at frequencies of 12.1 Hz to 12.0 Hz, which corresponds to Reynolds numbers of 11,000 to 13,000 and reduced frequencies of 1.5 to 1.25. The flow was captured in a planar measurement region oriented perpendicular to the free stream direction and located in the wake approximately 150 mm downstream of the tail. For the initial interpretation of the measurements, the time-series PIV data were transformed into a quasi-three-dimensional representation of the wake structure using a passive convection model. For reference, measurements were also performed with a tethered FWMAV at comparable conditions. The first results, presented in the form of helicity isosurfaces, showed a good repeatability among flapping cycles and also qualitative agreement between the free-flight and the tethered cases, suggesting that the free-flight measurements were reliable and meaningful. 

While the obtained results hold promises for further processing as well as for future experiments with the current setup, the data quality could be further increased by certain improvements of the control approach as well as of the flow visualisation procedure itself. While our control approach was designed and tested primarily for low speeds (0.8 m/s to 1.2 m/s), the first tests revealed that the interpretation of data captured in a plane perpendicular to the free stream direction can be complicated, because the relatively strong induced flow of the flapping wings, aligned with the body that is pitched by $\approx 70^\circ$ to $\approx 50^\circ$, hits the measurement plane at considerable angles. This phenomena will diminish with increasing speed, which allowed to perform meaningful measurements at speeds of 2 m/s to 2.4 m/s. For even better results, the control approach should be revised, as the dynamics of the DelFly become much more challenging at these speeds. However, even with a further increased position accuracy, the stereoscopic PIV method limits the measurements to be conducted in the wake only, as measurements closer to or even around the wings would have to deal with laser reflections on the wings. 

To enable reliable measurements around the wings, but also meaningful measurements at lower speeds, we recommend using a true 3D visualisation method such as tomographic PIV \cite{Scarano2013} for future experiments. Standard tomo-PIV using conventional seeding is not feasible, however, for the measurement volume size and data acquisition rate required for the present experimental conditions. Recent developments have explored the potential of achieving large-scale tomographic measurements by using small (sub-millimetre) neutrally-buoyant helium-filled soap bubbles as tracer particles \cite{Kuhn2011}. Although several studies have indeed proven the feasibility of this approach, we decided not to employ this method for the first trials, because as a relatively new method it still has its own challenges, many of which are related to the soap bubbles used as seeding particles. They are being employed due to their high reflectivity, which is needed when the laser beam of finite power is expanded to illuminate larger volumes. However, the soap bubbles tend to stick to the FWMAV wing foils, which negatively affects the wing operation over time. For this reason, the exposure of the wings to the particles needs to be as limited as possible, which needs a specific measurement strategy to be used that minimises this effect.

The initial results presented here proved that the developed methodology provides a reliable and repeatable way of obtaining PIV data in free flight and that the data quality is comparable to what is usually achieved in a (traditional) tethered setting. Moreover, this new approach, employing flying robots instead of animals, enables to perform measurements not only in steady state, but also during arbitrary controlled and reproducible manoeuvres. It also allows for further systematic investigations of parameter changes such as wing span, aspect ratio, wing flexibility, etc., something that was not possible before in free flight.




\section*{Acknowledgements}
We thank Sarah Gluschitz for making the nice sketch of the experimental setup.




\begin{thebibliography}{61}
\providecommand{\natexlab}[1]{#1}
\providecommand{\url}[1]{\texttt{#1}}
\expandafter\ifx\csname urlstyle\endcsname\relax
  \providecommand{\doi}[1]{doi: #1}\else
  \providecommand{\doi}{doi: \begingroup \urlstyle{rm}\Url}\fi

\bibitem[Keennon et~al.(2012)Keennon, Klingebiel, Won, and
  Andriukov]{Keennon2012}
Matthew Keennon, Karl Klingebiel, Henry Won, and Alexander Andriukov.
\newblock {Development of the Nano Hummingbird: A Tailless Flapping Wing Micro
  Air Vehicle}.
\newblock \emph{AIAA paper 2012-0588}, pages 1--24, 2012.

\bibitem[Ma et~al.(2013)Ma, Chirarattananon, Fuller, and Wood]{Ma2013}
Kevin~Y Ma, Pakpong Chirarattananon, Sawyer~B Fuller, and Robert~J Wood.
\newblock {Controlled Flight of a Biologically Inspired, Insect-Scale Robot}.
\newblock \emph{Science}, 340:\penalty0 603--607, 2013.

\bibitem[Gaissert et~al.(2014)Gaissert, Mugrauer, Mugrauer, Jebens, Jebens, and
  Knubben]{Gaissert2014}
Nina Gaissert, Rainer Mugrauer, G{\"{u}}nter Mugrauer, Agalya Jebens, Kristof
  Jebens, and Elias~Maria Knubben.
\newblock {Inventing a micro aerial vehicle inspired by the mechanics of
  dragonfly flight}.
\newblock In \emph{Lecture Notes in Computer Science (including subseries
  Lecture Notes in Artificial Intelligence and Lecture Notes in
  Bioinformatics)}, volume 8069 LNAI, pages 90--100. Springer Berlin
  Heidelberg, 2014.
\newblock \doi{10.1007/978-3-662-43645-5{\_}11}.

\bibitem[de~Croon et~al.(2009)de~Croon, de~Clerq, Ruijsink, Remes, and {De
  Wagter}]{deCroon2009}
G.~C. H.~E. de~Croon, K~M~E de~Clerq, R~Ruijsink, B.~D.~W. Remes, and C.~{De
  Wagter}.
\newblock {Design, aerodynamics, and vision-based control of the DelFly}.
\newblock \emph{International Journal of Micro Air Vehicles}, 1\penalty0
  (2):\penalty0 71--97, 2009.
\newblock \doi{10.1260/175682909789498288}.

\bibitem[Ristroph et~al.(2013)Ristroph, Ristroph, Morozova, Bergou, Chang,
  Guckenheimer, Wang, and Cohen]{Ristroph2013}
Leif Ristroph, Gunnar Ristroph, Svetlana Morozova, Attila~J. Bergou, Song
  Chang, John Guckenheimer, Z.~Jane Wang, and Itai Cohen.
\newblock {Active and passive stabilization of body pitch in insect flight.}
\newblock \emph{Journal of the Royal Society, Interface / the Royal Society},
  10\penalty0 (85):\penalty0 20130237, 2013.
\newblock \doi{10.1098/rsif.2013.0237}.

\bibitem[Muijres et~al.(2014{\natexlab{a}})Muijres, Elzinga, Melis, and
  Dickinson]{Muijres2014}
Florian~T. Muijres, Michael~J Elzinga, Johan~M Melis, and Michael~H. Dickinson.
\newblock {Flies Evade Looming Targets by Executing Rapid Visually Directed
  Banked Turns}.
\newblock \emph{Science}, 344\penalty0 (6180):\penalty0 172--177, apr
  2014{\natexlab{a}}.
\newblock \doi{10.1126/science.1248955}.

\bibitem[Sholtis et~al.(2015)Sholtis, Shelton, and Hedrick]{Sholtis2015}
Katherine~M. Sholtis, Ryan~M. Shelton, and Tyson~L. Hedrick.
\newblock {Field Flight Dynamics of Hummingbirds during Territory Encroachment
  and Defense}.
\newblock \emph{PLOS ONE}, 10\penalty0 (6):\penalty0 e0125659, jun 2015.
\newblock ISSN 1932-6203.
\newblock \doi{10.1371/journal.pone.0125659}.

\bibitem[Bomphrey et~al.(2016)Bomphrey, Nakata, Henningsson, and
  Lin]{Bomphrey2016}
Richard~J. Bomphrey, Toshiyuki Nakata, Per Henningsson, and Huai-Ti Lin.
\newblock {Flight of the dragonflies and damselflies}.
\newblock \emph{Philosophical Transactions of the Royal Society of London B:
  Biological Sciences}, 371\penalty0 (1704), 2016.

\bibitem[Dickinson et~al.(1999)Dickinson, Lehmann, and Sane]{Dickinson1999}
Michael~H. Dickinson, Fritz-Olaf Lehmann, and Sanjay~P Sane.
\newblock {Wing Rotation and the Aerodynamic Basis of Insect Flight}.
\newblock \emph{Science}, 284\penalty0 (5422):\penalty0 1954--1960, 1999.
\newblock \doi{10.1126/science.284.5422.1954}.

\bibitem[Sane and Dickinson(2002)]{Sane2002}
Sanjay~P Sane and Michael~H. Dickinson.
\newblock {The aerodynamic effects of wing rotation and a revised quasi-steady
  model of flapping flight}.
\newblock \emph{Journal of Experimental Biology}, 205:\penalty0 1087--1096,
  2002.

\bibitem[Berman and Wang(2007)]{Berman2007}
Gordon~J Berman and Z~Jane Wang.
\newblock {Energy-minimizing kinematics in hovering insect flight}.
\newblock \emph{Journal of Fluid Mechanics}, 582:\penalty0 153--168, 2007.
\newblock \doi{10.1017/S0022112007006209}.

\bibitem[Armanini et~al.(2016)Armanini, Caetano, de~Croon, de~Visser, and
  Mulder]{Armanini2016}
S~F Armanini, J~V Caetano, G~C H~E de~Croon, C~C de~Visser, and M~Mulder.
\newblock {Quasi-steady aerodynamic model of clap-and-fling flapping MAV and
  validation using free-flight data}.
\newblock \emph{Bioinspiration {\&} Biomimetics}, 11\penalty0 (4):\penalty0
  046002, jun 2016.
\newblock \doi{10.1088/1748-3190/11/4/046002}.

\bibitem[Nakata et~al.(2011)Nakata, Liu, Tanaka, Nishihashi, Wang, and
  Sato]{Nakata2011}
T~Nakata, H~Liu, Y~Tanaka, N~Nishihashi, X~Wang, and A~Sato.
\newblock {Aerodynamics of a bio-inspired flexible flapping-wing micro air
  vehicle.}
\newblock \emph{Bioinspiration {\&} biomimetics}, 6\penalty0 (4):\penalty0
  045002, dec 2011.
\newblock \doi{10.1088/1748-3182/6/4/045002}.

\bibitem[Deng et~al.(2015)Deng, Xiao, Percin, van Oudheusden, Bijl, and
  Remes]{Deng2015}
Shuanghou Deng, Tianhang Xiao, Mustafa Percin, Bas van Oudheusden, Hester Bijl,
  and Bart Remes.
\newblock {Numerical simulation of an X-wing flapping wing MAV by means of a
  deforming overset grid method}.
\newblock In \emph{22nd AIAA Computational Fluid Dynamics Conference}, Reston,
  Virginia, jun 2015. American Institute of Aeronautics and Astronautics.
\newblock \doi{10.2514/6.2015-2615}.

\bibitem[Tay et~al.(2016)Tay, de~Baar, Percin, Deng, and van
  Oudheusden]{Tay2016}
Wee~Beng Tay, J.H.S. de~Baar, Mustafa Percin, Shuanghou Deng, and Bas~W. van
  Oudheusden.
\newblock {Numerical simulation of a flapping wing MAV based on wing
  deformation capture analysis}.
\newblock In \emph{34th AIAA Applied Aerodynamics Conference}, Reston,
  Virginia, jun 2016. American Institute of Aeronautics and Astronautics.
\newblock \doi{10.2514/6.2016-3552}.

\bibitem[Deng(2016)]{Deng2016thesis}
Shuanghou Deng.
\newblock \emph{{Aerodynamics of flapping-wing Micro-Air-Vehicle: An integrated
  experimental and numerical study}}.
\newblock PhD thesis, Delft University of Technology, 2016.

\bibitem[Nakata and Liu(2011)]{Nakata2011flexible}
Toshiyuki Nakata and Hao Liu.
\newblock {Aerodynamic performance of a hovering hawkmoth with flexible wings:
  a computational approach}.
\newblock \emph{Proceedings of the Royal Society of London B: Biological
  Sciences}, 2011.

\bibitem[Thomas et~al.(2004)Thomas, Taylor, Srygley, Nudds, and
  Bomphrey]{Thomas2004}
Adrian L.~R. Thomas, Graham~K. Taylor, Robert~B. Srygley, Robert~L. Nudds, and
  Richard~J. Bomphrey.
\newblock {Dragonfly flight: free-flight and tethered flow visualizations
  reveal a diverse array of unsteady lift-generating mechanisms, controlled
  primarily via angle of attack}.
\newblock \emph{Journal of Experimental Biology}, 207\penalty0 (24), 2004.

\bibitem[Ellington et~al.(1996)Ellington, van~den Berg, Willmott, and
  Thomas]{Ellington1996}
Charles~P Ellington, Coen van~den Berg, Alexander~P Willmott, and Adrian L~R
  Thomas.
\newblock {Leading-edge vortices in insect flight}.
\newblock \emph{Nature}, 384:\penalty0 626--630, 1996.
\newblock \doi{10.1038/384626a0}.

\bibitem[Willmott et~al.(1997)Willmott, Ellington, and Thomas]{Willmott1997c}
Alexander~P Willmott, Charles~P Ellington, and Adrian L~R Thomas.
\newblock {Flow visualization and unsteady aerodynamics in the flight of the
  hawkmoth, Manduca sexta}.
\newblock \emph{Philos Trans R Soc Lond B Biol Sci.}, 352:\penalty0 303--316,
  1997.
\newblock \doi{10.1098/rstb.1997.0022}.

\bibitem[Fuchiwaki et~al.(2013)Fuchiwaki, Kuroki, Tanaka, and
  Tababa]{Fuchiwaki2013}
Masaki Fuchiwaki, Taichi Kuroki, Kazuhiro Tanaka, and Takahide Tababa.
\newblock {Dynamic behavior of the vortex ring formed on a butterfly wing}.
\newblock \emph{Experiments in Fluids}, 54\penalty0 (1):\penalty0 1450, jan
  2013.
\newblock ISSN 0723-4864.
\newblock \doi{10.1007/s00348-012-1450-x}.

\bibitem[Henningsson et~al.(2015)Henningsson, Michaelis, Nakata, Schanz,
  Geisler, Schr{\"{o}}der, and Bomphrey]{henningssonEtAl2015}
Per Henningsson, Dirk Michaelis, Toshiyuki Nakata, Daniel Schanz, Reinhard
  Geisler, Andreas Schr{\"{o}}der, and Richard~J Bomphrey.
\newblock {The complex aerodynamic footprint of desert locusts revealed by
  large-volume tomographic particle image velocimetry.}
\newblock \emph{Journal of the Royal Society Interface}, 12\penalty0 (108),
  2015.

\bibitem[Warrick et~al.(2005)Warrick, Tobalske, and Powers]{Warrick2005}
Douglas~R Warrick, Bret~W Tobalske, and Donald~R Powers.
\newblock {Aerodynamics of the Hovering Hummingbird}.
\newblock \emph{Nature}, 435:\penalty0 1094--1097, 2005.
\newblock \doi{10.1038/nature03647}.

\bibitem[Warrick et~al.(2009)Warrick, Tobalske, and Powers]{Warrick2009}
Douglas~R Warrick, Bret~W Tobalske, and Donald~R Powers.
\newblock {Lift production in the hovering hummingbird}.
\newblock \emph{Proceedings of the Royal Society B}, 276:\penalty0 3747--3752,
  2009.
\newblock \doi{10.1098/rspb.2009.1003}.

\bibitem[Altshuler et~al.(2009)Altshuler, Princevac, Pan, and
  Lozano]{Altshuler2009}
Douglas~L Altshuler, Marko Princevac, Hansheng Pan, and Jesse Lozano.
\newblock {Wake patterns of the wings and tail of hovering hummingbirds}.
\newblock \emph{Experiments in Fluids}, 45\penalty0 (5):\penalty0 835--846,
  2009.
\newblock \doi{10.1007/s00348-008-0602-5}.

\bibitem[Pournazeri et~al.(2013)Pournazeri, Segre, Princevac, and
  Altshuler]{Pournazeri2013}
Sam Pournazeri, Paolo~S. Segre, Marko Princevac, and Douglas~L. Altshuler.
\newblock {Hummingbirds generate bilateral vortex loops during hovering:
  evidence from flow visualization}.
\newblock \emph{Experiments in Fluids}, 54\penalty0 (1):\penalty0 1439, jan
  2013.
\newblock \doi{10.1007/s00348-012-1439-5}.

\bibitem[Bomphrey et~al.(2009)Bomphrey, Taylor, and Thomas]{Bomphrey2009}
Richard~James Bomphrey, Graham~K. Taylor, and Adrian L.~R. Thomas.
\newblock {Smoke visualization of free-flying bumblebees indicates independent
  leading-edge vortices on each wing pair}.
\newblock \emph{Experiments in Fluids}, 46\penalty0 (5):\penalty0 811--821, may
  2009.
\newblock \doi{10.1007/s00348-009-0631-8}.

\bibitem[Hedenstr{\"{o}}m et~al.(2009)Hedenstr{\"{o}}m, Muijres, von Busse,
  Johansson, Winter, and Spedding]{Hedenstrom2009}
Anders Hedenstr{\"{o}}m, F.~T. Muijres, R.~von Busse, L.~C. Johansson,
  Y.~Winter, and G.~R. Spedding.
\newblock {High-speed stereo DPIV measurement of wakes of two bat species
  flying freely in a wind tunnel}.
\newblock \emph{Experiments in Fluids}, 46\penalty0 (5):\penalty0 923--932, may
  2009.
\newblock \doi{10.1007/s00348-009-0634-5}.

\bibitem[Muijres et~al.(2014{\natexlab{b}})Muijres, {Christoffer Johansson},
  Winter, and Hedenstr{\"{o}}m]{Muijres2014bats}
Florian~T. Muijres, L~{Christoffer Johansson}, York Winter, and Anders
  Hedenstr{\"{o}}m.
\newblock {Leading edge vortices in lesser long-nosed bats occurring at slow
  but not fast flight speeds.}
\newblock \emph{Bioinspiration {\&} biomimetics}, 9\penalty0 (2):\penalty0
  025006, may 2014{\natexlab{b}}.
\newblock \doi{10.1088/1748-3182/9/2/025006}.

\bibitem[Johansson et~al.(2013)Johansson, Engel, Kelber, {Klein Heerenbrink},
  and Hedenstr{\"{o}}m]{Johansson2013}
L~Christoffer Johansson, Sophia Engel, Almut Kelber, Marco {Klein Heerenbrink},
  and Anders Hedenstr{\"{o}}m.
\newblock {Multiple leading edge vortices of unexpected strength in freely
  flying hawkmoth}.
\newblock \emph{Scientific Reports}, 3:\penalty0 3264, nov 2013.

\bibitem[Ortega-Jimenez et~al.(2014)Ortega-Jimenez, Sapir, Wolf, Variano, and
  Dudley]{Ortega-Jimenez2014}
Victor~M Ortega-Jimenez, Nir Sapir, Marta Wolf, Evan~A Variano, and Robert
  Dudley.
\newblock {Into turbulent air: size-dependent effects of von K{\'{a}}rm{\'{a}}n
  vortex streets on hummingbird flight kinematics and energetics.}
\newblock \emph{Proceedings. Biological sciences / The Royal Society},
  281\penalty0 (1783):\penalty0 20140180, may 2014.
\newblock \doi{10.1098/rspb.2014.0180}.

\bibitem[Berg and Ellington(1997{\natexlab{a}})]{VanDenBerg1997a}
Coen Van~Den Berg and Charles~P Ellington.
\newblock {The vortex wake of a `hovering' model hawkmoth}.
\newblock \emph{Philos Trans R Soc Lond B Biol Sci.}, 352:\penalty0 317--328,
  1997{\natexlab{a}}.
\newblock \doi{10.1098/rstb.1997.0023}.

\bibitem[Berg and Ellington(1997{\natexlab{b}})]{VanDenBerg1997b}
Coen Van~Den Berg and Charles~P Ellington.
\newblock {The three-dimensional leading-edge vortex of a `hovering' model
  hawkmoth}.
\newblock \emph{Philos Trans R Soc Lond B Biol Sci.}, 352:\penalty0 329--340,
  1997{\natexlab{b}}.
\newblock \doi{10.1098/rstb.1997.0024}.

\bibitem[Birch et~al.(2004)Birch, Dickson, and Dickinson]{Birch2004}
James~M Birch, William~B. Dickson, and Michael~H. Dickinson.
\newblock {Force production and flow structure of the leading edge vortex on
  flapping wings at high and low Reynolds numbers}.
\newblock \emph{Journal of Experimental Biology}, 207:\penalty0 1063--1072,
  2004.
\newblock \doi{10.1242/jeb.00848}.

\bibitem[Lehmann et~al.(2005)Lehmann, Sane, and Dickinson]{Lehmann2005}
Fritz-Olaf Lehmann, Sanjay~P Sane, and Michael~H. Dickinson.
\newblock {The aerodynamic effects of wing-wing interaction in flapping insect
  wings}.
\newblock \emph{Journal of Experimental Biology}, 208:\penalty0 3075--3092,
  2005.
\newblock \doi{doi:10.1242/jeb.01744}.

\bibitem[Truong et~al.(2013)Truong, Kim, Kim, Park, Yoon, and Byun]{Truong2013}
Van~Tien Truong, Jihoon Kim, Min~Jun Kim, Hoon~Cheol Park, Kwang~Joon Yoon, and
  Doyoung Byun.
\newblock {Flow structures around a flapping wing considering ground effect}.
\newblock \emph{Experiments in Fluids}, 54\penalty0 (7):\penalty0 1575, jul
  2013.
\newblock \doi{10.1007/s00348-013-1575-6}.

\bibitem[Cheng et~al.(2013)Cheng, Roll, Liu, Troolin, and Deng]{Cheng2013}
Bo~Cheng, Jesse Roll, Yun Liu, Daniel~R. Troolin, and Xinyan Deng.
\newblock {Three-dimensional vortex wake structure of flapping wings in
  hovering flight}.
\newblock \emph{Journal of The Royal Society Interface}, 11\penalty0 (91),
  2013.

\bibitem[Zheng et~al.(2016)Zheng, Wu, and Tang]{Zheng2016}
Yingying Zheng, Yanhua Wu, and Hui Tang.
\newblock {A time-resolved PIV study on the force dynamics of flexible tandem
  wings in hovering flight}.
\newblock \emph{Journal of Fluids and Structures}, 62:\penalty0 65--85, 2016.
\newblock \doi{10.1016/j.jfluidstructs.2015.12.008}.

\bibitem[Ren et~al.(2013)Ren, Wu, and Huang]{Ren2013}
H~Ren, Y~Wu, and P~G Huang.
\newblock {Visualization and characterization of near-wake flow fields of a
  flapping-wing micro air vehicle using PIV}.
\newblock \emph{Journal of Visualization}, 16\penalty0 (1):\penalty0 75--83,
  2013.
\newblock \doi{10.1007/s12650-012-0152-z}.

\bibitem[Percin et~al.(2014)Percin, van Oudheusden, Eisma, and
  Remes]{Percin2014}
Mustafa Percin, B.~W. van Oudheusden, H.~E. Eisma, and B.~D.~W. Remes.
\newblock {Three-dimensional vortex wake structure of a flapping-wing micro
  aerial vehicle in forward flight configuration}.
\newblock \emph{Experiments in Fluids}, 55, 2014.
\newblock \doi{10.1007/s00348-014-1806-5}.

\bibitem[Deng and van Oudheusden(2016)]{Deng2016}
Shuanghou Deng and Bas van Oudheusden.
\newblock {Wake structure visualization of a flapping-wing Micro-Air-Vehicle in
  forward flight}.
\newblock \emph{Aerospace Science and Technology}, 50:\penalty0 204--211, 2016.
\newblock \doi{10.1016/j.ast.2016.01.003}.

\bibitem[Deng et~al.(2016)Deng, Percin, and van
  Oudheusden]{Deng2016experimentalInvestigation}
Shuanghou Deng, Mustafa Percin, and Bas van Oudheusden.
\newblock {Experimental Investigation of Aerodynamics of Flapping-Wing
  Micro-Air-Vehicle by Force and Flow-Field Measurements}.
\newblock \emph{AIAA Journal}, 54\penalty0 (2):\penalty0 588--602, 2016.

\bibitem[Wiken(2015)]{wiken2015}
James~Neil Wiken.
\newblock \emph{{Analysis of a Quadrotor in Forward Flight}}.
\newblock Master's thesis, Massachusetts Institute of Technology, Cambridge,
  US-MA, 2015.

\bibitem[Nowak(2010)]{nowak2010}
Jan Nowak.
\newblock \emph{{Windkanal Freiflugmessungen zur Bestimmung flugmechanischer
  Kenngr\"{o}ssen}}.
\newblock Phd thesis, RWTH Aachen University, Aachen, DE, 2010.

\bibitem[{De Wagter} et~al.(2013{\natexlab{a}}){De Wagter}, Koopmans, de~Croon,
  Remes, and Ruijsink]{deWagterEtAl2013}
Christophe {De Wagter}, Andries Koopmans, Guido C H~E de~Croon, Bart D~W Remes,
  and Rick Ruijsink.
\newblock {Autonomous Wind Tunnel Free-Flight of a Flapping Wing MAV}.
\newblock In \emph{Advances in Aerospace Guidance, Navigation and Control},
  pages 603--621. Springer, Berlin, DE, 2013{\natexlab{a}}.

\bibitem[Caetano et~al.(2015)Caetano, Percin, van Oudheusden, Remes, {De
  Wagter}, de~Croon, and de~Visser]{Caetano2015}
J~V Caetano, M~Percin, B~.~W. van Oudheusden, B.~D.~W. Remes, C.~{De Wagter},
  G.~C. H.~E. de~Croon, and C.~C. de~Visser.
\newblock {Error analysis and assessment of unsteady forces acting on a
  flapping wing micro air vehicle: free flight versus wind-tunnel experimental
  methods.}
\newblock \emph{Bioinspiration {\&} biomimetics}, 10\penalty0 (5):\penalty0
  056004, oct 2015.
\newblock \doi{10.1088/1748-3190/10/5/056004}.

\bibitem[de~Croon et~al.(2016)de~Croon, Percin, Remes, Ruijsink, and {De
  Wagter}]{deCroon2016delflyBook}
G.~C. H.~E. de~Croon, Mustafa Percin, B.~D.~W. Remes, Rick Ruijsink, and C.~{De
  Wagter}.
\newblock \emph{{The DelFly - Design, Aerodynamics, and Artificial Intelligence
  of a Flapping Wing Robot}}.
\newblock Springer Netherlands, 2016.
\newblock \doi{10.1007/978-94-017-9208-0}.

\bibitem[{De Clercq} et~al.(2010){De Clercq}, de~Kat, Remes, van Oudheusden,
  and Bijl]{DeClercq2010}
Kristien~M.E. {De Clercq}, Roeland de~Kat, B.~D.~W. Remes, Bas~W. van
  Oudheusden, and Hester Bijl.
\newblock {Aerodynamic Experiments on DelFly II: Unsteady Lift Enhancement}.
\newblock \emph{International Journal of Micro Air Vehicles}, 1\penalty0
  (4):\penalty0 255--262, 2010.
\newblock \doi{10.1260/175682909790291465}.

\bibitem[Koopmans et~al.(2015)Koopmans, Tijmons, {De Wagter}, and
  de~Croon]{Koopmans2015}
J.A. Koopmans, S.~Tijmons, C.~{De Wagter}, and G.~C. H.~E. de~Croon.
\newblock {Passively Stable Flapping Flight From Hover to Fast Forward Through
  Shift in Wing Position}.
\newblock \emph{International Journal of Micro Air Vehicles}, 7\penalty0 (4),
  jan 2015.

\bibitem[Remes et~al.(2014)Remes, Esden-Tempski, {Van Tienen}, Smeur, {De
  Wagter}, and de~Croon]{Remes2014}
B.~D.~W. Remes, P.~Esden-Tempski, F.~{Van Tienen}, E.~Smeur, C.~{De Wagter},
  and G.~C. H.~E. de~Croon.
\newblock {Lisa-S 2.8g autopilot for GPS-based flight of MAVs}.
\newblock In \emph{IMAV 2014: International Micro Air Vehicle Conference and
  Competition 2014, Delft, The Netherlands, August 12-15, 2014}, pages
  280--285, aug 2014.

\bibitem[Pap(2016)]{Paparazzi}
{Paparazzi UAV}, 2016.
\newblock URL \url{http://wiki.paparazziuav.org/wiki/Main_Page}.

\bibitem[Muijres et~al.(2011)Muijres, Johansson, Winter, and
  Hedenstr{\"{o}}m]{Muijres2011}
Florian~T Muijres, L~Christoffer Johansson, York Winter, and Anders
  Hedenstr{\"{o}}m.
\newblock {Comparative aerodynamic performance of flapping flight in two bat
  species using time-resolved wake visualization.}
\newblock \emph{Journal of the Royal Society, Interface / the Royal Society},
  8\penalty0 (63):\penalty0 1418--28, oct 2011.
\newblock \doi{10.1098/rsif.2011.0015}.

\bibitem[Henningsson et~al.(2011)Henningsson, Muijres, and
  Hedenstr{\"{o}}m]{Henningsson2011}
P.~Henningsson, F.~T. Muijres, and A.~Hedenstr{\"{o}}m.
\newblock {Time-resolved vortex wake of a common swift flying over a range of
  flight speeds}.
\newblock \emph{Journal of The Royal Society Interface}, 8\penalty0 (59), 2011.

\bibitem[Muijres et~al.(2012)Muijres, Bowlin, Johansson, and
  Hedenstr{\"{o}}m]{Muijres2012flycatchers}
Florian~T Muijres, Melissa~S Bowlin, L~Christoffer Johansson, and Anders
  Hedenstr{\"{o}}m.
\newblock {Vortex wake, downwash distribution, aerodynamic performance and
  wingbeat kinematics in slow-flying pied flycatchers.}
\newblock \emph{Journal of the Royal Society, Interface / the Royal Society},
  9\penalty0 (67):\penalty0 292--303, feb 2012.
\newblock \doi{10.1098/rsif.2011.0238}.

\bibitem[Theys et~al.(2016)Theys, {De Vos}, and {De Schutter}]{Theys2016}
B.~Theys, G.~{De Vos}, and J.~{De Schutter}.
\newblock {A control approach for transitioning VTOL UAVs with continuously
  varying transition angle and controlled by differential thrust}.
\newblock In \emph{2016 International Conference on Unmanned Aircraft Systems
  (ICUAS)}, pages 118--125. IEEE, jun 2016.
\newblock \doi{10.1109/ICUAS.2016.7502519}.

\bibitem[{De Wagter} et~al.(2013{\natexlab{b}}){De Wagter}, Dokter, de~Croon,
  and Remes]{DeWagter2013}
Christophe {De Wagter}, Dirk Dokter, Guido C H~E de~Croon, and Bart D~W Remes.
\newblock {Multi-lifting-device uav autonomous flight at any transition
  percentage}.
\newblock In \emph{Proceedings of EuroGNC}, 2013{\natexlab{b}}.

\bibitem[Karasek et~al.(2016)Karasek, Koopmans, Armanini, Remes, and
  de~Croon]{Karasek2016}
Matej Karasek, Andries~Jan Koopmans, Sophie~F Armanini, Bart D~W Remes, and
  Guido C H~E de~Croon.
\newblock {Free flight force estimation of a 23.5 g flapping wing MAV using an
  on-board IMU}.
\newblock In \emph{The 2016 IEEE/RSJ International Conference on Intelligent
  Robots and Systems (IROS 2016), Daejeon, Korea, 9-14 October 2016}, 2016.

\bibitem[Cunis et~al.(2016)Cunis, Karasek, and de~Croon]{Cunis2016}
Torbj{\o}rn Cunis, Matej Karasek, and Guido C H~E de~Croon.
\newblock {Precision Position Control of the DelFly II Flapping-wing Micro Air
  Vehicle in a Wind-tunnel}.
\newblock In \emph{The International Micro Air Vehicle Conference and
  Competition 2016 (IMAV 2016), Beijing, China, October 17-21}, 2016.

\bibitem[Degani et~al.(1990)Degani, Seginer, and Levy]{Degani1990}
David Degani, Arnan Seginer, and Yuval Levy.
\newblock {Graphical visualization of vortical flows by means of helicity}.
\newblock \emph{AIAA Journal}, 28\penalty0 (8):\penalty0 1347--1352, aug 1990.
\newblock \doi{10.2514/3.25224}.

\bibitem[Scarano(2013)]{Scarano2013}
F~Scarano.
\newblock {Tomographic PIV: principles and practice}.
\newblock \emph{Measurement Science and Technology}, 24\penalty0 (1):\penalty0
  012001, 2013.
\newblock \doi{10.1088/0957-0233/24/1/012001}.

\bibitem[K{\"{u}}hn et~al.(2011)K{\"{u}}hn, Ehrenfried, Bosbach, and
  Wagner]{Kuhn2011}
Matthias K{\"{u}}hn, Klaus Ehrenfried, Johannes Bosbach, and Claus Wagner.
\newblock {Large-scale tomographic particle image velocimetry using
  helium-filled soap bubbles}.
\newblock \emph{Experiments in Fluids}, 50\penalty0 (4):\penalty0 929--948, apr
  2011.
\newblock \doi{10.1007/s00348-010-0947-4}.

\end{thebibliography}

\end{document}